\documentclass[reprint,amsmath,amssymb,aps,prd]{revtex4-2}

\usepackage{dcolumn}
\usepackage{bm}
\usepackage{subcaption}
\usepackage{hyperref}
\usepackage[pdftex]{graphicx}
\usepackage{amsmath, amssymb, amsthm, braket,mathrsfs}
\usepackage{bbold}
\usepackage{bm}

\usepackage{tikz}
\usepackage{tikz-3dplot}
\usetikzlibrary{calc}
\usepackage[T1]{fontenc}

\DeclareMathOperator{\Tr}{Tr}

\DeclareMathOperator{\SO}{SO}
\DeclareMathOperator{\SU}{SU}

\DeclareMathOperator{\su}{\mathfrak{su}}

\DeclareMathOperator{\Ad}{Ad}
\DeclareMathOperator{\diag}{diag}

\DeclareMathOperator{\SPD}{SPD}

\DeclareMathOperator{\Id}{Id}

\begin{document}

\preprint{APS/123-QED}

\title{Generalized skyrmion crystals with applications to neutron stars}
\author{Paul Leask}\email{mmpnl@leeds.ac.uk (Corresponding Author)}
\affiliation{School of Mathematics, University of Leeds, Leeds, LS2 9JT, England, UK}
\author{Miguel Huidobro}\email{miguel.huidobro.garcia@usc.es}
\affiliation{Departamento de F\'isica de Part\'iculas, Universidad de Santiago de Compostela and Instituto Galego de F\'isica de Altas Enerxias (IGFAE), Santiago de Compostela, E-15782, Spain}
\author{Andrzej Wereszczynski}\email{andrzej.wereszczynski@uj.edu.pl}
\affiliation{Institute of Physics, Jagiellonian University, Lojasiewicza 11, Krak\'{o}w, Poland}

\date{\today}

\begin{abstract}
In this article we study properties of isospin asymmetric nuclear matter in the generalized $\mathcal{L}_{0246}$-Skyrme model.
This is achieved by canonically quantizing the isospin collective degrees of freedom of the recently found skyrmion multi-wall crystal.
We obtain, for the first time, an equation of state from the Skyrme model which interpolates between infinite isospin asymmetric nuclear matter and finite isospin symmetric atomic nuclei.
This enables us to describe neutron stars with crusts within the Skyrme framework.
Furthermore, we observe that the symmetry energy tends to a constant value at zero density, which can be identified with the asymmetry coefficient in the semi-empirical mass formula for atomic nuclei.
The symmetry energy also reveals a cusp in its structure below the nuclear saturation point $n_0$ at $n_*\sim 3n_0/4$.
This cusp density point $n_*$ can be interpreted as the nuclear density whereby the infinite crystalline multi-wall configuration undergoes a phase transition to a finite isolated multi-wall configuration.
Both of these observations are observed to be generic features of skyrmion crystals that tend asymptotically to somewhat isolated skyrmion configurations in the zero density limit.
We find that the resulting neutron stars from our study agree quite well with recent NICER/LIGO observational data.
\end{abstract}

\keywords{Topological solitons; Skyrme model; skyrmions}
\maketitle



\section{Introduction}

The Skyrme model \cite{Skyrme_1961} offers a unique, unified framework in which one can study baryonic matter at all scales - from single baryons and atomic nuclei to infinite nuclear matter which, after coupling the model to gravity, gives rise to neutron stars \cite{Manton_2022}.
All of this emerges from an elegantly simple Lagrangian containing a limited number of terms and, in consequence, a few free coupling constants, where the fundamental degrees of freedom (d.o.f.) are the lightest mesons disguised into a matrix valued field.
In the minimal version, which is used in this work, they are pions forming an $\SU(2)$-valued field.
The attractiveness of this approach originates not only in a very small number of parameters but also in the manifestation of baryons.
Namely, they are realized as non-perturbative excitations of the mesonic field, that is, as topological solitons, called skyrmions.
Importantly, the topological degree of skyrmions has been identified with the baryon charge in a rigorous way \cite{Witten_1983_1,Witten_1983_2}.  

The Skyrme model has been very extensively studied in the context of nucleons \cite{Adkins_1983, Adkins_1984}, and light atomic nuclei \cite{Braaten_1986, Braaten_1988, Leese_1995, Barnes_1997, Carson_1991, Battye_2005, Battye_2006, Manko_2007, Battye_2007, Battye_2009} with many spectacular results.
In particular, let us mention the description of the ground and Hoyle states in $^{12}C$ \cite{Lau_2014} and excitation bands of $^{16}O$ \cite{Halcrow_2017, Manton_2019} as well as the emergence of $\alpha$-cluster structure \cite{Naya_2018} which is expected for not too heavy atomic nuclei.
This recent progress to large extent relies on an improved quantization procedure where, contrary to the usual rigid-rotor approach, one takes into account both the zero modes and the softest massive vibrations \cite{Halcrow_2016}.
Also, the long standing problem of binding energies has found a resolution by inclusion of additional terms \cite{Adam_2010, Adam:2013wya, Gillard_2015, Gillard_2017, Gudnason:2016tiz} or additional mesonic degrees of freedom \cite{Sutcliffe:2010et, Sutcliffe:2011ig, Naya:2018mpt, Gudnason_2020}, both physically well motivated.
Finally, it is now clear how to extract nuclear forces from the Skyrme model \cite{Halcrow:2020gbm, Harland_2021}, which ultimately may provided a much better contact with more traditional nuclear models. 

Obviously, a natural field of application of the Skyrme framework is nuclear matter and neutron stars.
However, a correct description of this regime is still a serious challenge for the solitonic Skyrme model. 
 
The problem of \textit{infinite nuclear matter at non-zero density} can be approached if one considers the model on a finite volume unit cell with periodic boundary conditions \cite{Klebanov_1985}, which results in an infinite but periodic Skyrme crystal.
Varying the volume of the unit cell (while keeping the baryon number fixed) allows one to study skyrmionic matter at finite densities and, inter alia, to obtain an equation of state (EoS).
Taking the advantage of the Tolman--Oppenheimer--Volkoff construction, one obtains neutron stars.
This approach meets some difficulties both at the mathematical and physical level. 

First of all, in the traditional approach to determining the EoS, the geometry of the unit cell was \textit{fixed to be cubic} and the unit cell volume was varied by homothety about the cell center.
This rendered the energy minimizing crystalline solutions to inherit the symmetry group of their corresponding initial configuration. In a consequence one obtained not true energy minimizers but solutions with imposed geometrical structure. 

Nevertheless, various crystal solutions were constructed \cite{Klebanov_1985, Goldhaber_1987, Kugler_1988, Kugler_1989, Castillejo_1989}.
This led to a \textit{conjecture} that, at moderate and large densities, the global energy minimizer should be very well approximated by the simple cubic crystal of half-skyrmions \cite{Kugler_1988,Castillejo_1989} ($\textup{SC}_{1/2}$) \footnote{
In previous works, this $\textup{SC}_{1/2}$ configuration  of half-skyrmions was also referred to as a face centered cubic (FCC) crystal.
This FCC designation is due to the initial configuration being that of $B=1$ hedgehog skyrmions arranged on an FCC lattice, however it relaxes to that of the simple cubic crystal of half-skyrmions.
For larger enough unit cell volume, this $\textup{SC}_{1/2}$ crystal undergoes a phase transition back to the FCC crystal of hedgehogs.
However, we choose to label this crystal the $\textup{SC}_{1/2}$ crystal.}.
At even larger densities a transition to the body centered cubic crystal of half-skyrmions ($\textup{BCC}_{1/2}$) is observed \cite{Goldhaber_1987, Adam:2021gbm}.

Secondly, even if one accepts this constrained approach, three important physical issues have been reported: 
\begin{enumerate}
\item High density issue: the EoS is too soft, giving rise to neutron stars that are too light.
\item Low density issue: the presence of a minimum at saturation in the EoS yields negative pressure, which represents a thermodynamically unstable phase at low density. 
\item Saturation density issue: nuclear binding energies are too large, which in turn means the compression modulus is too large.
\end{enumerate}

The too softness of the standard Skyrme model EoS was found an elegant resolution by extension to the \textit{generalized $\mathcal{L}_{0246}$-Skyrme model} where the so-called sextic term has been included.
Indeed, this component of the $\mathcal{L}_{0246}$-Skyrme Lagrangian was essential not only to significantly increases the value of the maximal mass of neutron star (from $1.7 M_{\odot}$ \cite{Nelmes:2011zz} to above $2M_{\odot}$ \cite{Adam_2020}), but also to render nuclear matter more like a perfect fluid, especially at higher densities, which corresponds very well to the standard picture of a (super-)fluid core of neutron star.
These results are deeply anchored in the mathematical properties of the sextic term.
Namely, if treated together with the (pion mass) potential term, the corresponding energy-momentum tensor has a perfect fluid form \cite{Naya_2014}.
In addition it enjoys a volume preserving diffeomorphism symmetry which means that the energy of a solution is degenerate up all deformations which do not change its volume \cite{Adam_2010}.
On the contrary, deformations that change the volume are strongly penalized as the corresponding EoS has a maximally stiff form \cite{Naya_2015,Vazquez_2015}.
This agrees with a physical interpretation of the sextic term as a part of the action which effectively arises after integration of $\omega$-mesons.
Indeed, EoS' obtained in the Walecka model at large densities tend to maximally stiff EoS' due to the $\omega$-meson repulsion. 

At low density the situation is much less clear due to the appearance of \textit{thermodynamically unstable regions}.
This is directly related to the use of the fixed geometry approach mentioned above.
For each fixed classical crystal solution (which, in a natural way, is identified with symmetric nuclear matter), the energy $E$ per unit cell possesses a minimum for a certain volume $V_*$, which may be consistently identified with the saturation point.
Obviously, for $V>V_*$, the solution is thermodynamically unstable as it formally corresponds to negative pressure.
However, taking into account the isospin quantum corrections and some further contributions, the classical minimum should disappear, thereby providing a thermodynamically stable description even in the low density regime.
This periodic crystal was then expected to be replaced by \textit{non-homogeneous solutions} in this regime \cite{SilvaLobo_2009, SilvaLobo_2011, Park:2019bmi, Adam_2022}.
Here \cite{sym15040899} for example, a crystal of $\alpha$-particles and $B=32$ skyrmions have been considered.
Although these configurations lowered the classical energy per cell, they did not cure the instability issue \cite{Wereszczynski_2022}.
In conclusion, the Skyrme model provided an EoS, but only above the saturation point, leaving the low density regime rather completely unexplored. This is obviously a serious problem of the Skyrme framework as the lower density regime is typically identified with the crust of neutron star.
In fact in these densities many geometrically involved phases are expected, see, e.g., lasagna or pasta phases. 

Of course, at low densities the isospin and, especially, Coulomb effects should be taken into account.
Although it is conceptually clear how it should be done (e.g., taking into account semi-classical quantization of the isospin d.o.f. \cite{Baskerville_1996}), and some interesting outcomes have been recently reported \cite{Wereszczynski_2022}, the results where obtained for the \textit{fixed} $\textup{SC}_{1/2}$ crystal of half-skyrmions, which is \emph{not} the global minimizer at any density \cite{Leask_2023}. 

In any case, a thermodynamically stable phase(s) at low density is the first necessary step in approaching of the problem of the crust and nuclear pasta phases within the solitonic Skyrme model. 

Finally, there is a very famous problem of the computation of the compression modulus.
It is the quadratic term in expansion of EoS of infinite symmetric nuclear matter at the saturation density $n_0$.
The widely accept value, based on the vibrating frequency extracted from the Isoscalar Giant Monopole Resonance, is $K_0=240 \pm 20$MeV.
Depending on a version of the Skyrme model with massive pions, as well as on a particular choice of the coupling constants, one gets value a few times bigger than expected.
Namely, $K_0 \sim 1350-2300$ MeV.
More importantly, this result was derived for fixed geometry crystals.
It has been therefore advocated that the actual value should be lower if a non-homogeneous solution would be the true minimizer.
This again brings us back to necessity of solving the generalized $\mathcal{L}_{0246}$-Skyrme model at finite density without any geometric constraints. 

It is the aim of the current paper to simultaneously face all these problems by constructing generalized skyrmion crystals at finite density without any symmetry assumptions on the skyrmion nor its fundamental period lattice.
This is possible due to the recently developed method of obtaining crystalline solutions by not only considering the variation of the Skyrme field $\varphi: \mathbb{R}^3/\Lambda \rightarrow \SU(2)$ but also by allowing non-cubic variations of the unit cell period lattice $\Lambda$  \cite{Leask_2023}.

The main idea is the identification between all $3$-tori $(\mathbb{R}^3/\Lambda,d)$, with $d$ the Euclidean metric, and the unit $3$-torus $\mathbb{T}^3=\mathbb{R}^3/\mathbb{Z}^3$, where $\mathbb{T}^3$ is equipped with the flat pullback metric $g=F^*d$ via a diffeomorphism $F: \mathbb{T}^3 \rightarrow \mathbb{R}^3/\Lambda$.
Varying the metric $g$ on $\mathbb{T}^3$ is equivalent to considering variations of the period lattice $\Lambda$.
It is convenient to think of the metric $g$ as a constant symmetric-positive-definite matrix $(g_{ij})$.
Then one can address this variational problem by identifying the gradient of the energy with respect to the metric $(g_{ij})$ with the stress-energy tensor $(S_{ij})$ of the field $\varphi$.
Auckly and Kapitanski \cite{Kapitanski_2003} showed that, for fixed metric $g$, the energy functional $E(\varphi,g)$ attains a minimum.
In \cite{Leask_2023} they proved that, for fixed field configuration $\varphi$, any critical metric $g$ of the energy functional $E(\varphi,g)$ is in fact a unique local minimum.
Hence the period lattice $\Lambda$, for which the Skyrme field $\varphi$ has minimum energy, is unique (up to automorphism).
This means that the resulting periodic crystalline configuration is indeed a true energy minimizer with respect to both variations of the period lattice $\Lambda$ and the Skyrme field $\varphi$.
This slightly improves on the known crystalline solutions at medium and large densities, i.e. above the nuclear saturation point, but it has a tremendous impact on the low density regime where non-homogeneous solutions are expected to exist. 

Here, we apply this method to the generalized $\mathcal{L}_{0246}$-Skyrme model and obtain the lattice ground state of the generalized model at all densities, that is, above and below the nuclear saturation point $n_0$.
In our model, infinite nuclear matter is not necessarily treated as being homogeneous.
At saturation $n_0$, it appears as an almost homogeneous multi-wall configuration with near cubic symmetry.
At low densities ($n<n_0$) then it is considered inhomogeneous, with distinct somewhat isolated multi-wall configurations present.
Whereas, at high densities ($n>n_0$), e.g. in the core, it appears even more homogeneous and as a simple cubic crystal of fractional half-skyrmions, i.e. it merges with the $\textup{SC}_{1/2}$ crystal.
This allows us, for the first time, to obtain an EoS of the skyrmionic matter which interpolates between low and high density regimes.
We use this EoS, with an addition of the \textit{isospin quantum contribution} and with the assumption of \textit{$\beta$-equilibrium}, and investigate its usefulness and consequences for nuclear physics. 


\section{Skyrme crystals and phases of skyrmion matter}
\label{sec: Skyrme crystals and phases of skyrmion matter}


\subsection{The generalized $\mathcal{L}_{0246}$-Skyrme model}
\label{subsec: The Skyrme model}

The generalized $\mathcal{L}_{0246}$-Skyrme model consists of a single scalar field $\varphi: \Sigma \rightarrow \SU(2)$ where spacetime is given by the $(3+1)$-dimensional Lorentzian manifold $\Sigma = \mathbb{R} \times M$ with the product metric $g = -\textup{d}t^2 + h$, and $(M,h)$ is an oriented $3$-dimensional Riemannian manifold with Riemannian metric $h$.
Let us introduce oriented local coordinates $(x^0,x^1,x^2,x^3)$ on the domain $\Sigma$ and let $\{ \partial_0, \partial_1, \partial_2, \partial_3 \}$ be a local basis for the tangent space $T_x \Sigma$ at $x \in \Sigma$, where we have denoted $\partial/\partial x^\mu \equiv \partial_\mu$.
We equip $\su(2)$ with the $\Ad(\SU(2))$ invariant inner product $(X,Y)_{\su(2)} = \frac{1}{2} \Tr(X^\dagger Y)$.
Let $\Omega \in \Omega^2(\SU(2)) \otimes \su(2)$ be an $\su(2)$-valued two-form on $\SU(2)$ and $\omega \in \Omega^1(\SU(2)) \otimes \su(2)$ be the left Maurer-Cartan form.
Then, for any left invariant vector fields $X,Y \in T_{\varphi(x)}\SU(2)$, where $x \in \Sigma$, we define
\begin{equation}
    \Omega(X,Y) = \left[ \omega(X), \omega(Y) \right],
\end{equation}
where $[\cdot, \cdot]: \su(2) \times \su(2) \rightarrow \su(2)$ is the usual Lie bracket.
The left Maurer-Cartan form $\omega$ defines the $\su(2)$-valued \textit{left current}
\begin{equation}
    L_\mu := \omega_\varphi (\partial_\mu \varphi) = \varphi^\dagger \partial_\mu \varphi.
\end{equation}
Let us write the pullback as $\Omega_{\mu\nu}=\varphi^*\Omega(\partial_\mu,\partial_\nu)$.
Then the curvature can be expressed in terms of the $\su(2)$-valued left current as
\begin{align}
    \Omega_{\mu\nu} = \left[ L_\mu, L_\nu \right].
\label{eq: Skyrme model - Curvature}
\end{align}

Consider the trivial foliation of spacetime $\Sigma = \mathbb{R} \times M$ into spacelike hypersurfaces $M$ and let $M$ be compact and without boundary.
This is the case if, for example, $M$ is a $3$-torus or the usual vacuum boundary condition $\varphi(x\rightarrow \infty) = \mathbb{I}_2$ is imposed on $M=\mathbb{R}^3$.
Then Hopf's degree theorem ensures that such mappings $\varphi: M \rightarrow \SU(2)\cong S^3$, for $M=\mathbb{R}^3 \cup \{\infty\} \cong S^3$ and $\mathbb{T}^3$, are characterized by a homotopy invariant: the topological degree $B \in \mathbb{Z}$, since $\pi_3(S^3) = H_3(\mathbb{T}^3) = \mathbb{Z}$.
This topological degree is identified with the physical baryon number upon quantization, so we often to refer to $B$ as the baryon number, which may be computed using
\begin{equation}
\label{eq: Baryon number}
    B = \int_M \textup{d}^3x \sqrt{-g} \, \mathcal{B}^0,
\end{equation}
where the topological current is given by
\begin{equation}
\label{eq: Topological current}
    \mathcal{B}^\mu = \frac{1}{24 \pi^2 \sqrt{-g}} \epsilon^{\mu \nu \rho \sigma} \Tr(L_\nu L_\rho L_\sigma).
\end{equation}

We consider the generalization of the massive Skyrme Lagrangian which yields an $\omega$-meson-like repulsion on short distances, while also allowing the quartic Skyrme term to describe scalar meson effects.
This generalized Skyrme Lagrangian is composed of four terms and is given by
\begin{equation}
\label{eq: Generalized Lagrangian}
    \mathcal{L}_{0246} = \mathcal{L}_0 + \mathcal{L}_2 + \mathcal{L}_4 + \mathcal{L}_6,
\end{equation}
where the the index $i$ denotes the degree of each term as a polynomial in spatial derivatives.
The four terms appearing in the energy functional are the potential, Dirichlet, Skyrme and sextic terms, respectively.
It is conventional to label the models by terms used in the energy functional, e.g. the generalized model is labeled $\mathcal{L}_{0246}$, the standard massive model is denoted $\mathcal{L}_{024}$, the massless Skyrme model $\mathcal{L}_{24}$ and the BPS model $\mathcal{L}_{06}$.
The first term is the potential which provides a mass for the pionic fields,
\begin{equation}
    \mathcal{L}_0 =  - \frac{c_0}{8\hbar^3}F_\pi^2 m_\pi^2 \Tr\left( \mathbb{I}_2 - \varphi \right).
\end{equation}
The Dirichlet, or kinetic, term is given by
\begin{equation}
    \mathcal{L}_2 = c_2\frac{F_\pi^2}{16\hbar}g^{\mu\nu} \Tr(L_\mu L_\nu)
\end{equation}
and the Skyrme term, corresponding to the four pion interaction, is
\begin{equation}
    \mathcal{L}_4 =  \frac{c_4\hbar}{8e^2}g^{\mu\alpha}g^{\nu\beta} \Tr\left( \left[L_\mu, L_\nu\right] \left[L_\alpha, L_\beta\right] \right).
\end{equation}
Finally, we include the sextic term, defined by \cite{Jackson_1985}
\begin{equation}
    \mathcal{L}_6 =  -\pi^4 \lambda^2 g^{\mu\nu} \mathcal{B}_\mu \mathcal{B}_\nu,
\end{equation}
where $\mathcal{B}^\mu$ is the topological Chern--Simons current defined in \eqref{eq: Topological current}.
The $c_i$ are coupling constants and, for the usual $\mathcal{L}_{0246}$-Skyrme model, take the values $c_0=c_2=1$ and $c_4=1/4$.
The pion mass is fixed to take its physical value of $m_\pi=140\,\textup{MeV}$.
So, the free parameters of the model are the pion decay constant $F_\pi$,  the dimensionless Skyrme parameter $e$, and $\lambda$ which is related to the mass $m_\omega$ and coupling constant $g_\omega$ of the $\omega$ meson via $\lambda^2 = g_\omega^2 \hbar^3/(2\pi^4 m_\omega^2)$ \cite{Wereszczynski_2015}.
The reduced Planck constant is $\hbar=197.33$ MeV fm. 
Throughout we will use the values
\begin{equation}
    F_\pi=122\,\textup{MeV}, \quad e=4.54, \quad \lambda^2=1\,\textup{MeV\,fm}^3.
\label{eq: Coupling constants}
\end{equation}

Qualitatively, the parameters \eqref{eq: Coupling constants} don't have much affect on the ground state configuration.
However, quantitatively this is not true.
We fit the parameters of the model to give us approximately the binding energy at saturation and the nuclear density, while also allowing the symmetry energy and the pion decay constant not to deviate too much from their experimental values.
Other studies have done similar fittings to, e.g., the symmetry energy, but there is always a trade-off where if you fix one parameter accurately then another physical quantities will suffer in consequence.
In other studies \cite{Wereszczynski_2022}, the symmetry energy and saturation energy can be fitted correctly, but the saturation density can not also be simultaneously fitted correctly.
That is the caveat of using the Skyrme model alone to model nuclear matter.
For example, in our model, the symmetry energy at saturation is lower than expected but accurately predicts the asymmetry coefficient in the SEMF.
If the model is tuned to give the correct symmetry energy value at saturation then the asymmetry coefficient would be off. 
For a more general review of the quantitative effects of the free parameters on a ground state configuration, see \cite{Wereszczynski_2022,sym15040899}.

We are interested in static solutions and adopt the usual Skyrme units of length and energy.
The classical energy scale is $\tilde{E}=F_\pi/4e$ (MeV) and the length scale is $\tilde{L}=2\hbar/e F_\pi$ (fm).
Thus the quantum energy scale is defined by $\tilde{\hbar}= 2e^2$.
In these dimensionless Skyrme units, the rescaled pion mass for our studies is
\begin{equation}
    m = \frac{2 m_\pi}{F_\pi e}
\end{equation}
and the dimensionless sextic coupling constant is
\begin{equation}
    c_6 = \frac{\pi^4 \lambda^2 e^4 F_\pi^2}{2\hbar^3}.
\end{equation}
It will prove useful throughout to introduce the Hilbert energy-momentum tensor (in dimensionless Skyrme units):
\begin{align}
    T_{\mu\nu} = \, & -c_2 \Tr(L_\mu L_\nu) - c_4 g^{\alpha\beta}\Tr([L_\mu,L_\alpha][L_\nu,L_\beta]) \nonumber \\
    \, & + 2c_6 \mathcal{B}_\mu \mathcal{B}_\nu + g_{\mu\nu}\mathcal{L}_{0246}.
\end{align}
The static energy functional can be obtained from the timelike part of the energy-momentum tensor, $T_{00} = \mathcal{E}_\textup{stat} + \mathcal{E}_\textup{kin}$, and is given by
\begin{equation}
    M_B(\varphi,g) = \int_M \textup{d}^3x \sqrt{-g} \, \mathcal{E}_\textup{stat},
\label{eq: Skyrme model - Static energy}
\end{equation}
where
\begin{align}
    \mathcal{E}_\textup{stat} = \, & c_0 m^2 \Tr\left( \mathbb{I}_2 - \varphi \right) -\frac{c_2}{2}g^{ij} \Tr(L_i L_j) \nonumber \\
    \, & - \frac{c_4}{4}g^{ia}g^{jb} \Tr\left( [L_i, L_j] [L_a, L_b] \right) \nonumber \\
    \, & + c_6 \frac{\epsilon^{ijk}\epsilon^{abc} }{(24 \pi^2 \sqrt{-g})^2} \Tr(L_i L_j L_k) \Tr(L_a L_b L_c).
\end{align}

A field configuration $\varphi$ which minimizes the static energy functional \eqref{eq: Skyrme model - Static energy}, for some choice of domain metric $g$, is referred to as a skyrmion and the static energy $M_B$ is often interpreted as the classical mass of the skyrmion.
The associated Euler--Lagrange field equations can be approximately solved by discretizing the static energy \eqref{eq: Skyrme model - Static energy} and employing a 4th order central finite-difference method.
This is carried out using the quaternionic formulation detailed below.
We can then regard the static energy as a function $M_B:\mathcal{C} \rightarrow \mathbb{R}$, where the discretised configuration space is the manifold $\mathcal{C}=(S^3)^{N_1\,N_2\,N_3} \subset \mathbb{R}^{4\,N_1\,N_2\,N_3}$.
To solve the Euler--Lagrange field equations we use arrested Newton flow: an accelerated gradient descent method with flow arresting, with some appropriate initial configuration.
That is, we are solving the system of 2nd order ODEs
\begin{equation}
    \ddot{\varphi} = -\frac{\delta \mathcal{E}_\textup{stat}}{\delta \varphi}, \quad \varphi(0)=\varphi_0,
\label{eq: Numerical minimisation procedure - 2nd order ODEs}
\end{equation}
with initial velocity $\dot{\varphi}(0)=0$.
Setting $\psi:=\dot{\varphi}$ as the velocity with $\psi(0)=\dot{\varphi}(0)=0$ reduces the problem to a coupled system of 1st order ODEs.
We implement a 4th order Runge--Kutta method to solve this coupled system.
In general, the initial configuration $\varphi_0$ is not a minimizer and so it swaps its potential energy for kinetic energy as it evolves.
During the evolution we check to see if the energy is increasing.
If the energy is indeed increasing, we take out all the kinetic energy in the system by setting $\psi(t)=\dot{\varphi}(t)=0$ and restart the flow (this is the arresting criteria).
Naturally the field will relax to a local, or global, minimum in some potential well.
The evolution then terminates when every component of the energy gradient $\frac{\delta M_B}{\delta \varphi}$ is zero within some specified tolerance, e.g. $\textup{tol}=10^{-5}$.


\subsection{Metric independent integral formulation}
\label{subsec: Metric independent integral formulation}

For numerical purposes, it is convenient to utilize the quaternionic representation of the target group $\SU(2)$, which is topologically isomorphic to $S^3$.
Let us parameterize the unit quaternion $\varphi \in \mathbb{H}$ by the mesonic fields $(\varphi^0,\varphi^1,\varphi^2,\varphi^3)$:
\begin{equation}
    \SU(2) \ni \begin{pmatrix}
        \varphi^0 + i\varphi^3 & i\varphi^1 + \varphi^2 \\
        i\varphi^1 - \varphi^2 & \varphi^0 - i\varphi^3
    \end{pmatrix} \leftrightarrow (\varphi^0,\varphi^1,\varphi^2,\varphi^3) \in S^3,
\label{eq: Quaternion representation}
\end{equation}
with the unitary condition $\sigma^2 + \vec{\pi}\cdot\vec{\pi}=1$, where $\vec{\pi}=(\varphi^1,\varphi^2,\varphi^3)$ is normally identified with the triplet of pion fields and $\sigma=\varphi^0$ with the $\sigma$-field.
Then the Maurer-Cartan left current can be expressed as the vector quaternion:
\begin{align}
    L_i = \, & - i L^a_i \tau^a, \nonumber \\
    L^a_i = \, & \epsilon^{abc} \partial_i\varphi^b \varphi^c + \partial_i\varphi^0 \varphi^a - \partial_i\varphi^a \varphi^0,
\end{align}
where $\tau^a$ are the isospin Pauli matrices and, similarly, the curvature in the quaternionic representation is given by
\begin{align}
    \Omega_{ij} = \, & -2 i \Omega^a_{ij} \tau^a, \nonumber \\ 
    \Omega^a_{ij} = \, & \epsilon^{abc} \partial_i\varphi^b \partial_j\varphi^c + \partial_i\varphi^0 \partial_j\varphi^a - \partial_i\varphi^a \partial_j\varphi^0.
\end{align}
From this we get the following contractions,
\begin{subequations}
    \begin{align}
        L^a_i L^a_j = \, & \partial_i\varphi^\mu \partial_j\varphi^\mu, \\
        \Omega^a_{ij} \Omega^a_{kl} = \, & \partial_i\varphi^\mu \partial_k\varphi^\mu \partial_j\varphi^\nu \partial_l\varphi^\nu - \partial_i\varphi^\mu \partial_l\varphi^\mu \partial_j\varphi^\nu \partial_k\varphi^\nu, \\
        L_i^a \Omega_{jk}^a = \, & -\epsilon_{\mu\nu\alpha\beta}\varphi^\mu \partial_i\varphi^\nu \partial_j\varphi^\alpha \partial_k\varphi^\beta.
    \end{align}
\end{subequations}
The baryon number density in contraction form is
\begin{align}
    \mathcal{B}^0 = \, & \frac{1}{12\pi^2\sqrt{-g}} \epsilon^{ijk} L_i^a \Omega_{jk}^a.
\end{align}
For numerical simulations involving the minimization of the energy functional with respect to variations of the metric, it will be convenient to define the metric independent integrals:
\begin{subequations}
    \begin{align}
    \label{eq: Metric independent E0}
        W(\varphi) = \, & 2c_0 m^2 \int_{\mathbb{T}^3} \textup{d}^3x \, (1-\varphi^0), \\
    \label{eq: Metric independent E2}
        L_{ij}(\varphi) = \, & c_2 \int_{\mathbb{T}^3} \textup{d}^3x \, L^a_i L^a_j, \\
    \label{eq: Metric independent E4}
        \Omega_{ijkl}(\varphi) = \, & 2c_4 \int_{M} \textup{d}^3x \, \Omega_{ij}^a \Omega_{kl}^a, \\
    \label{eq: Metric independent E6}
        C(\varphi) = \, & c_6 \frac{\epsilon^{ijk}\epsilon^{lmn} }{(12\pi^2)^2} \int_{M} \textup{d}^3x \, L_i^a \Omega_{jk}^a L_l^b \Omega_{mn}^b.
    \end{align}
\end{subequations}
In terms of these metric independent integrals, the static energy can be compactly written as
\begin{equation}
\begin{split}
    M_B(\varphi,g) = \sqrt{-g} W(\varphi) + \sqrt{-g}g^{ij}L_{ij}(\varphi) \\
     + \sqrt{-g}g^{ik}g^{jl}\Omega_{ijkl}(\varphi) + \frac{C(\varphi)}{\sqrt{-g}}.
\end{split}
\end{equation}


\subsection{Skyrme crystals}
\label{subsec: Skyrme crystals}

Our aim is to study Skyrme fields $\varphi:\mathbb{R}^3\rightarrow\SU(2)$ that are periodic with respect to some $3$-dimensional period lattice $\Lambda$, i.e. we impose the condition $\varphi(x+X)=\varphi(x)$ for all $x\in\mathbb{R}^3$ and $X \in \Lambda$.
We can equivalently interpret the field as a map $\varphi: \mathbb{R}^3/\Lambda \rightarrow \SU(2)$, where $(\mathbb{R}^3/\Lambda,d)$ is a $3$-torus equipped with the standard Euclidean metric $d$.
In particular, we define a Skyrme crystal to be an energy minimizing map
\begin{equation}
    \varphi: \mathbb{R}^3/\Lambda_\diamond \rightarrow \SU(2), \quad \Lambda_\diamond = \left\{ \sum_{i=1}^3 n_i \vec{X}_i : n_i \in \mathbb{Z} \right\},
\end{equation}
where $\mathbb{R}^3/\Lambda_\diamond$ is some fixed $3$-torus such that the field $\varphi$ is also critical and stable with respect to variations of the lattice $\Lambda$ about $\Lambda_\diamond$.
The problem of determining Skyrme crystals was addressed by Harland \textit{et al.} \cite{Leask_2023}.
They prove that, for a fixed field configuration $\varphi$, there is a unique period lattice $\Lambda_\diamond$ (up to automorphism) that minimizes the static energy $M_B$.
Therefore, the problem of determining skyrmion crystals is one of finding critical points of the static energy functional \eqref{eq: Skyrme model - Static energy} with respect to variations of both the field $\varphi$ and the period lattice $\Lambda_\diamond$.

For massless $\mathcal{L}_{24}$-skyrmions, the period lattice can be determined explicitly.
However, only a numerical approach seems possible for generalized $\mathcal{L}_{0246}$-skyrmions.
For some initial period lattice $\Lambda_0$, the static energy can minimized with respect to variations of the period lattice using the method detailed in \S\ref{subsec: Numerical optimization of the lattice geometry}.
In tandem, with some appropriate initial field configuration $\varphi_0$, the static energy functional can also be minimized with respect to variations of the field by using arrested Newton flow (ANF), which is detailed in \S\ref{subsec: The Skyrme model}.

Skyrme crystals have been studied extensively in the literature, with it being previously accepted that the $\textup{SC}_{1/2}$ crystal found independently by Kugler \& Shtrikmann \cite{Kugler_1988} and Castillejo \textit{et al.} \cite{Castillejo_1989} is the minimum energy Skyrme crystal.
However, in the massless $\mathcal{L}_{24}$-Skyrme model, this $\textup{SC}_{1/2}$ crystal is just one point on an $\SO(4)$ orbit of solutions, i.e. it is not an isolated critical point and all of these solutions are all energy degenerate.
When the pion mass is turned on, there is no reason to expect these degenerate $\mathcal{L}_{24}$ critical points to extend to $\mathcal{L}_{0246}$ critical points upon perturbation.
However, there are four critical points which survive perturbation as argued by \cite{Leask_2023}.
These are the $\textup{SC}_{1/2}$, $\alpha$, chain and multi-wall crystals.
Each crystal has baryon number $B_\textup{cell}=4$ per unit cell, with three of the crystals having lower energy classically than the $\textup{SC}_{1/2}$ crystal for non-zero pion mass and non-cubic (trigonal) lattice geometry.

The $\textup{SC}_{1/2}$ crystal can be obtained from the Fourier series-like expansion of the fields as an initial configuration \cite{Castillejo_1989},
\begin{equation}
\label{eq: New crystals from old - Fourier}
    \varphi^0  = - c_1 c_2 c_3, \quad \varphi^1 = s_1 \sqrt{1-\frac{s_2^2}{2}-\frac{s_3^2}{2}+\frac{s_2^2 s_3^2}{3}},
\end{equation}
and cyclic, where $s_i=\sin(2\pi x^i/L)$ and $c_i=\cos(2\pi x^i/L)$.
From the $\textup{SC}_{1/2}$ crystal, the other three crystals can be constructed by applying a chiral $\SO(4)$ transformation $Q\in\SO(4)$, such that $\varphi=Q\varphi_{\textup{SC}_{1/2}}$, and minimizing the energy with respect to variations of the field and the lattice.
These chiral transformations $Q\in\SO(4)$ can be determined by considering a deformed energy functional on the moduli space of critical points of the Skyrme energy functional, and are found to be \cite{Leask_2023}
\begin{equation}
    \begin{split}
        Q \in \left\{
        \underbrace{\begin{pmatrix}
        (1,0,0,0) \\
        * \\
    \end{pmatrix}}_{Q_{\textup{SC}_{1/2}}},
    \underbrace{\begin{pmatrix}
        (0,-1,1,1)/\sqrt{3} \\
        * \\
    \end{pmatrix}}_{Q_\alpha}, \right. \\ \left.
    \underbrace{\begin{pmatrix}
        (0,0,0,1) \\
        * \\
    \end{pmatrix}}_{Q_\textup{multi-wall}}, 
    \underbrace{\begin{pmatrix}
        (0,0,1,1)/\sqrt{2} \\
        * \\
    \end{pmatrix}}_{Q_\textup{chain}}
    \right\}.
    \end{split}
\end{equation}
The other three rows of the chiral transformations $Q_\alpha$, $Q_\textup{multi-wall}$ and $Q_\textup{chain}$, labeled by the asterisk, can be obtained by using the Gram--Schmidt process.

Out of the four crystal configurations, the most of interest to astrophysicists are the $\alpha$-crystal, chain-crystal and the multi-wall crystal; these resemble non-uniform phases of nuclear matter, known as nuclear ``pasta''.
The iron rich crust of a neutron star could be modeled by $B=56$ chunks of $\alpha$-particle crystals, such as those modeled by Feist \textit{et al.} \cite{Feist_2013}, describing the ``gnocchi'' phase.
As we descend deeper towards the outer core, the pressure due to gravity increases and nuclei are squeezed together into long thin tubes of ``spaghetti''.
This spaghetti phase can be modeled using the chain-crystal.
Deeper still and the spaghetti flattens into parallel multi-walls, resembling ``lasagna'', of which the multi-wall crystal is reminiscent of.
Of course, for realistic applications the Coulomb interaction must be added.
This is because of the fact that different crust phases arise due to a balance between the strong and electrostatic forces.
Nevertheless, the Skyrme model has a built-in ability to model such phases. 

The multi-wall-crystal is the lowest energy solution at all baryon densities and also yields a lower compression modulus than the other three crystals.
This makes it an ideal candidate for nuclear matter and an equation of state (EoS) at high and low densities.
With $\varphi_0=Q_\textup{multi-wall}\varphi_{\textup{SC}_{1/2}}$ as an initial configuration and by considering fixed baryon density variations, as laid out in \S\ref{subsec: Phases of skyrmion matter}, the energy-volume curve can be computed and an EoS obtained.


\subsection{The stress-energy tensor}
\label{subsec: The stress-energy tensor}

To determine Skyrme crystal solutions, we identify every $3$-torus $(\mathbb{R}^3/\Lambda, d)$, equipped with the standard Euclidean metric $d$, with the unit $3$-torus $(\mathbb{T}^3,g)$ where $g$ is a Riemannian metric and $\mathbb{T}^3=\mathbb{R}^3/\mathbb{Z}^3$.
This metric $g$ on $\mathbb{T}^3$ is the pullback $g=F^* d$, with $g_{ij} = \vec{X}_i \cdot \vec{X}_j$, via the diffeomorphism 
\begin{align}
    F: & \, (\mathbb{T}^3, g) \rightarrow (\mathbb{R}^3/\Lambda, d), \nonumber \\
    & \, (x^1,x^2,x^3) \mapsto x^1\vec{X}_1 + x^2\vec{X}_2 + x^3\vec{X}_3.
\end{align}
Let the Skyrme field be the map $\varphi \circ F: \mathbb{T}^3 \rightarrow \SU(2)$.
We vary the metric $g_s$ on $\mathbb{T}^3$ with $g_0=F^* d$ which is equivalent to varying the lattice $\Lambda_s$ with $\Lambda_0=\Lambda$.
The energy minimized over variations $g_s$ of the domain metric is equivalent to determining the energy minimizing period lattice $\Lambda_\diamond$.

Now let the static Skyrme field be the smooth map $\varphi: \mathbb{T}^3 \rightarrow \SU(2)$.
Let $(x^1,x^2,x^3)$ be oriented local coordinates on $\mathbb{T}^3$ and $\{\partial_1,\partial_2,\partial_3\}$ be a local frame for the tangent space $T_x\mathbb{T}^3$ at $x \in \mathbb{T}^3$.
Let $g_s$ be a smooth one-parameter family of metrics on $\mathbb{T}^3$ with $g_0=F^* d$.
Set $\delta g = \partial_s g_s |_{s=0} \in \Gamma(\odot^2 T^*\mathbb{T}^3)$, a symmetric 2-covariant tensor field on $\mathbb{T}^3$.
Denote the inner product on the space of $2$-covariant tensor fields of the tangent space $T_x \mathbb{T}^3$ to $\mathbb{T}^3$ at $x \in \mathbb{T}^3$ by $\braket{\cdot,\cdot}$.
Then for any pair of symmetric bilinear forms $A,B$ we have
\begin{equation}
    \braket{A,B}_h = A_{ij} g^{jk} B_{kl} g^{li}.
\end{equation}
In particular, we have the following result:
\begin{equation}
    \Tr_g(A) = \braket{A,g}_g = g^{ij}A_{ij}.
\label{eq: Trace}
\end{equation}
Let us consider the rate of change of the energy of the Skyrme field $\varphi$ with respect to varying the domain metric $g$.
The first variation of the energy with respect to the variation $g(s)$ of the metric on $\mathbb{T}^3$ is given by
\begin{equation}
    \left.\frac{\textup{d} M_B(\varphi, g_s)}{\textup{d}s}\right|_{s=0} = \int_{\mathbb{T}^3} \textup{d}^3x \sqrt{g}  \braket{S(\varphi,g), \delta g}_g,
\end{equation}
where $S(\varphi,g) \in \Gamma(\odot^2 T^*\mathbb{T}^3)$ is a symmetric $2$-covariant tensor field on $\mathbb{T}^3$, known as the \textit{stress-energy tensor}, defined by
\begin{align}
    S_{ij} = \, & \frac{1}{2} \left[ c_0 m^2 \Tr(\Id - \varphi) - \frac{c_2}{2}g^{kl}\Tr(L_k L_l) \right. \nonumber \\
    \, & \left. - \frac{c_4}{4}g^{km}g^{ln} \Tr([L_k,L_l][L_m,L_n]) - c_6 (B_0)^2 \right] g_{ij} \nonumber \\
    \, & + \frac{c_2}{2} \Tr(L_i L_j) + \frac{c_4}{2}g^{kl}\Tr([L_i,L_k][L_j,L_l]).
\end{align}
This stress-energy tensor is related to the spatial part of the (static) energy-momentum tensor,
\begin{equation}
    S_{ij} = \frac{1}{\sqrt{g}}\frac{\delta (\sqrt{g}\mathcal{L}_{0246})}{\delta g^{ij}} = -\frac{1}{2}\,T_{ij}.
\end{equation}


\subsection{Numerical optimization of the lattice geometry}
\label{subsec: Numerical optimization of the lattice geometry}

Let us fix the field $\varphi: \mathbb{T}^3 \rightarrow \SU(2)$ and think of the energy $M_B$ as a function of the metric $g$ on $\mathbb{T}^3$.
That is, we define a map $E_\varphi: \SPD_3 \rightarrow \mathbb{R}$ such that $E_\varphi:=M_B(\left.\varphi\right|_\textup{fixed},g)$, where $\SPD_3$ is the space of symmetric positive-definite $3 \times 3$-matrices.
To minimize the energy functional $E_\varphi$ with respect to variations of the metric $g_s$, we use arrested Newton flow on $\SPD_3$.
The essence of the algorithm is as follows: we solve Newton's equations of motion for a particle on $\SPD_3$ with potential energy $E_\varphi$.
Now let $g_s$ be a smooth one-parameter curve in $\SPD_3$ with $g_0=F^*d$.
Explicitly, we are solving the system of 2nd order ODEs
\begin{equation}
    \left.\frac{\textup{d}^2}{\textup{d}s^2}\right|_{s=0} (g_{ij})_s = -\frac{\partial E_\varphi}{\partial g_{ij}} = - \int_{\mathbb{T}^3} \textup{d}^3x \sqrt{g} \, S_\varphi^{ij},
\label{eq: Metric ANF}
\end{equation}
with initial condition $(g_{ij})_0=\vec{X}_i \cdot \vec{X}_j$, and where $S_\varphi=S(g)$ is the stress-energy tensor for fixed field configuration $\varphi$.
Setting $\delta g_s = \partial_s g_s$ as the velocity with initial velocity $\delta g_0 = \left.\partial_{s}g_s\right|_{s=0} = 0$ reduces the problem to a coupled system of 1st order ODEs.
We implement a 4th order Runge--Kutta method to solve this coupled system.
The components of the stress-energy tensor for fixed field $\varphi$, given in the metric independent integral formulation, reads
\begin{equation}
    \begin{split}
        \int_{\mathbb{T}^3} S_\varphi^{ij} \textup{vol}_g = \frac{1}{2} g^{ij} \sqrt{g} \, W + \sqrt{g} \left( \frac{1}{2} g^{mn}g^{ij} 
        - g^{im}g^{jn} \right) L_{mn} \\ + \sqrt{g} \left( \frac{1}{2}g^{ij}g^{ln} - 2 g^{il}g^{jn} \right) g^{km} \Omega_{klmn} - \frac{1}{2} g^{ij} \frac{C}{\sqrt{g}}.
    \end{split}
\label{eq: Metric stress tensor}
\end{equation}

In general, the dimension of $\SPD_n$ is $\dim(\SPD_n)=n(n+1)/2$.
In our case, we are working with $\SPD_3$ and consider the energy as a function $E_\varphi: \SPD_3 \rightarrow \mathbb{R}$.
So we are implementing arrested Newton flow on a 6 dimensional manifold.
After each time step $t \mapsto t + \delta t$, we check to see if the energy is increasing.
If $E_\varphi(t + \delta t) > E_\varphi(t)$, we take out all the kinetic energy in the system by setting $\delta g(t + \delta t)=0$ and restart the flow.
The flow then terminates when every component of the stress-energy tensor $S_\varphi$ is zero to within a given tolerance (we have used $10^{-6}$).


\begin{figure}[t]
    \centering
    \begin{subfigure}[b]{0.38\textwidth}
        \includegraphics[width=\textwidth]{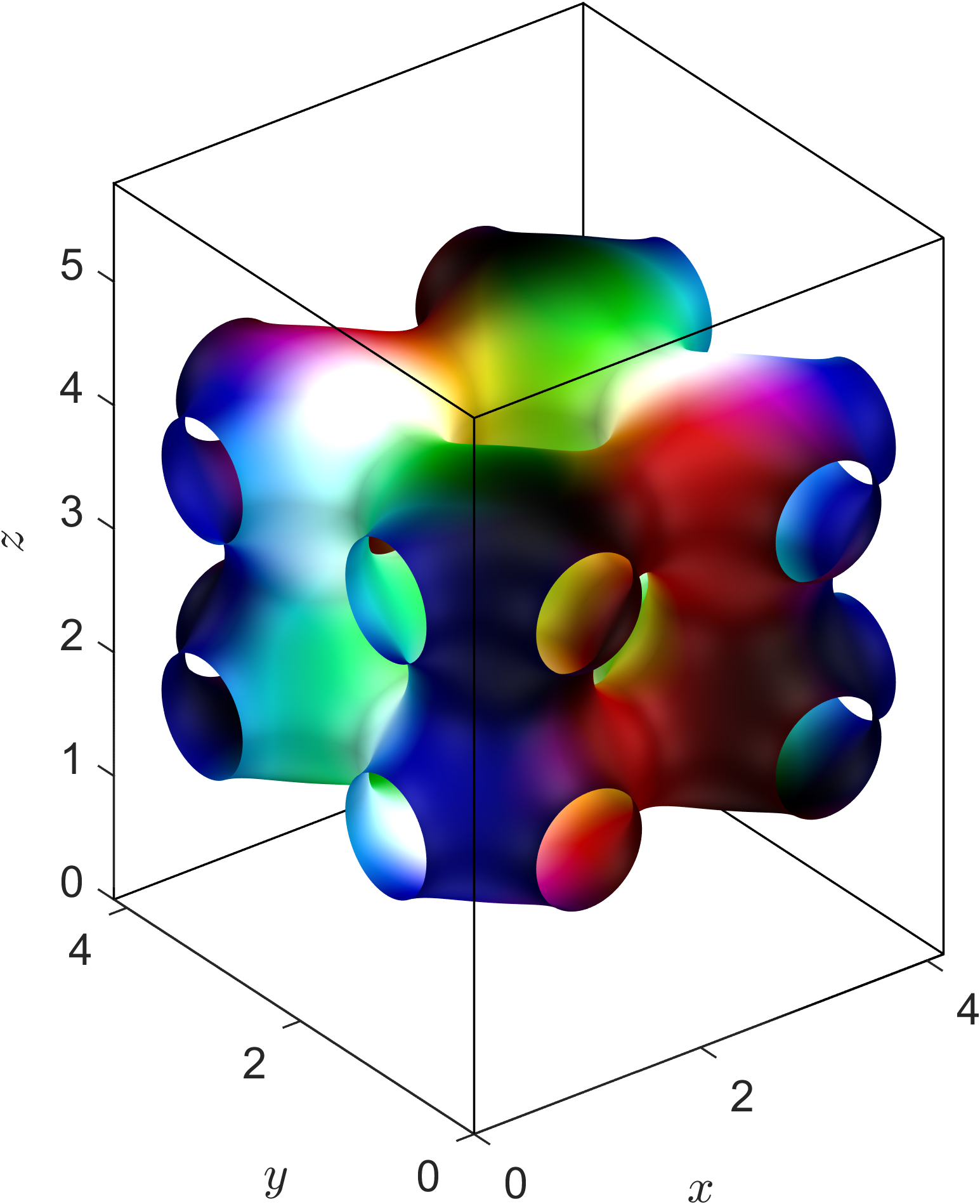}
        \caption{}
        \label{fig: BaryonDensity}
    \end{subfigure}
    \\
    \begin{subfigure}[b]{0.38\textwidth}
        \includegraphics[width=\textwidth]{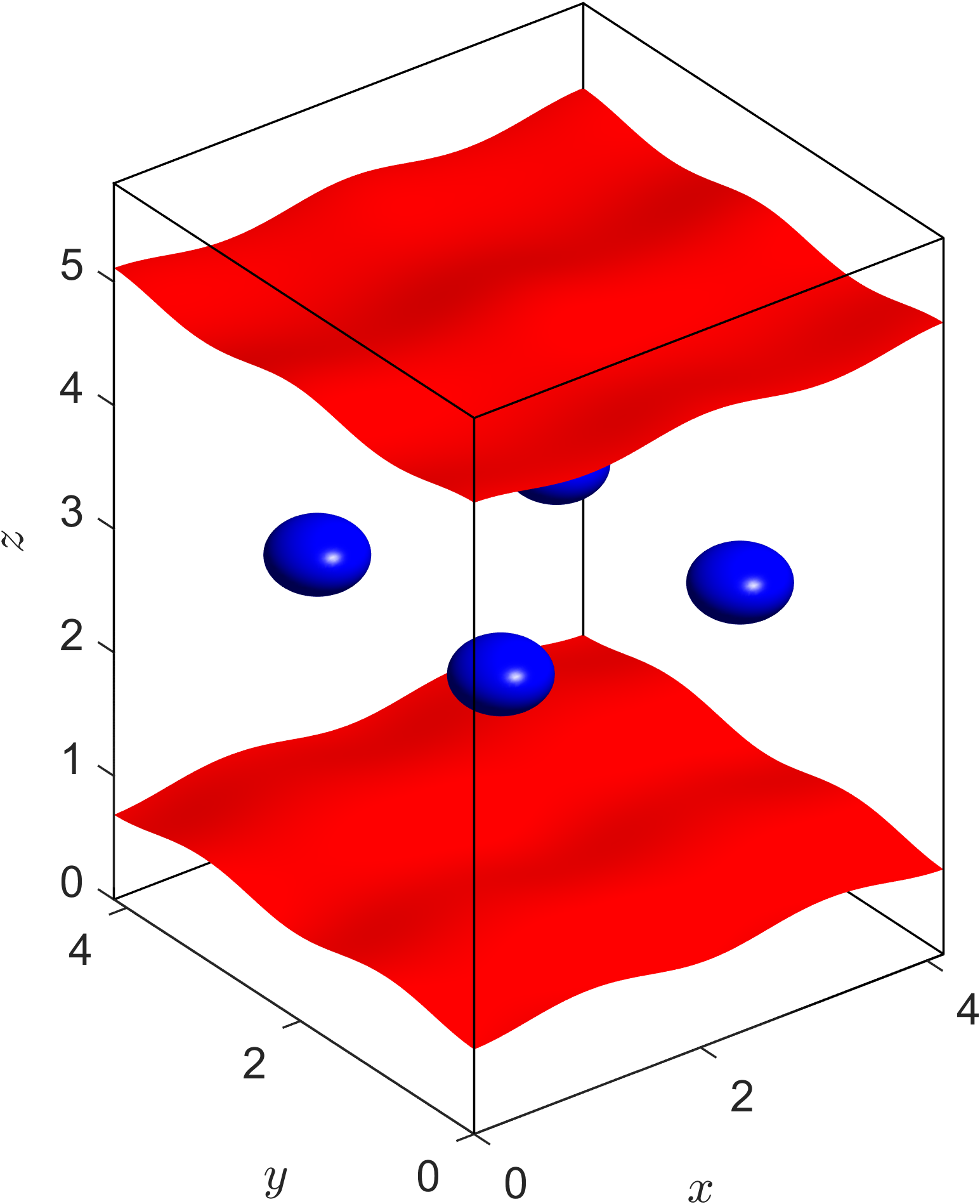}
        \caption{}
        \label{fig: SigmaPlots}
    \end{subfigure}
    \caption{$\mathcal{L}_{0246}$-Skyrme multi-wall crystal at a fixed baryon density $n_B<n_0$.
    The isobaryon density is depicted in (a) and isosurface plots of the $\sigma$ field, where the vacuum $(\sigma=+0.9)$ is colored red and the anti-vacuum $(\sigma=-0.9)$ blue, are shown in (b).}
    \label{fig: multi-wall crystal}
\end{figure}

\subsection{Phases of skyrmion matter}
\label{subsec: Phases of skyrmion matter}

Determining phases of nuclear matter and phase transitions in the Skyrme model is a difficult task, and is important if one wants to understand symmetric and asymmetric nuclear matter in high/low density regimes.
To study phases of matter at various densities, we consider fixed density variations of the energy functional, i.e. we allow the lattice to vary but keep its volume fixed.
Then the volume form $\textup{vol}_g$ is required to be invariant under variations $g_s$ of the metric, viz.
\begin{equation}
    \left.\frac{\textup{d}}{\textup{d}s}\right|_{s=0} \int_{\mathbb{T}^3} \textup{d}^3x \sqrt{g_s} = \frac{1}{2} \int_{\mathbb{T}^3} \textup{d}^3x \sqrt{g} g^{ij} \delta g_{ij} = 0.
\end{equation}
That is, $\delta g$ has to be an element of the space of traceless parallel symmetric bilinear forms $\mathscr{E}_0$.

In terms of the energy, we are dealing with a constrained minimization problem: minimize the energy functional for fixed field configuration $\varphi=\varphi|_{\textup{fixed}}$ subject to the constraint that $\det(g) = \textup{constant}$.
We can approach this using the method of Lagrange multipliers.
This leads to modifying the stress-energy tensor in \eqref{eq: Metric ANF} via the mapping
\begin{equation}
    S_\varphi \mapsto \tilde{S}_\varphi = S_\varphi - \frac{1}{3}\Tr_g(S_\varphi)\,g
\end{equation}
and our convergence criterion becomes $\max(\tilde{S}_\varphi) < \textup{tol}$.
Likewise, to ensure numerically that $\delta g$ is traceless, we need to project out the component of variation vector $\delta g$ parallel to the metric $g$ via the mapping
\begin{equation}
    \delta g \mapsto \delta g - \frac{1}{3} (g^{ij} \delta g_{ij}) \, g.
\end{equation}

By employing this process at various volumes it enables us to determine an energy-volume curve or, equivalently, an energy-density curve.
This is key to obtaining an EoS within our framework, as the EoS is directly related to the $E-V$ curve.

\subsection{The results}
\label{subsec: Resutls}

The first main result of this section is the observation that, as it is for the massive $\mathcal{L}_{024}$-Skyrme model, the multi-wall crystal is also the ground state crystalline solution for the generalized $\mathcal{L}_{0246}$-Skyrme model at all densities.
In the low density regime the solution clearly exhibits a two-layer structure, extending parallel to the $xy$-plane with the vacuum ($\sigma=1)$ occupying the regions above and below the multi-wall.
This can be seen in Fig.~\ref{fig: multi-wall crystal}.
Inside the multi-wall center the $\sigma$-field is approximately the anti-vacuum $(\sigma \approx -1)$.
Therefore, the multi-wall crystal is similar to that of a domain wall crystal, hence the name convention.
As the density increases, the regions occupied by the vacuum reduces and the non-cubic period lattice becomes more cubic, tending asymptotically to the $\textup{SC}_{1/2}$ crystal in the zero volume limit.
These are the true energy minimizers of the generalized $\mathcal{L}_{0246}$-Skyrme model at finite density under assumption that the baryon charge of the unit cell is four, $B_{\textup{cell}}=4$.

\begin{figure}[t]
    \centering
    \includegraphics[width=0.5\textwidth]{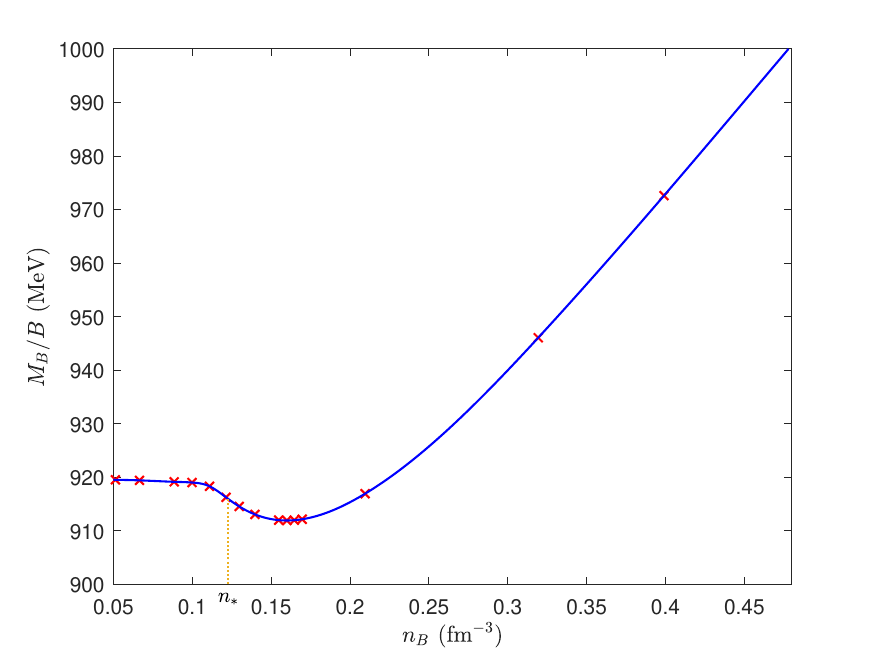}
    \caption{The classical static energy per baryon $M_B/B$ as a function of the nuclear density $n_B$.
    The nuclear density at which at the cusp in the symmetry energy appears is labeled by $n_*$.
    This corresponds to the density at which the infinite crystalline multi-wall solution begins transitioning to an isolated multi-wall configuration.}
    \label{fig: Plot_ClassicalEnergy_Density}
\end{figure}

In Fig. \ref{fig: Plot_ClassicalEnergy_Density} we plot the classical static energy per baryon $E=M_B/B$ of the multi-wall crystal as a function of the baryon density $n_B$.
This is interpreted as EoS of the symmetric nuclear matter since the classical Skyrme model does not distinguish between protons and neutrons.

Expansion of the energy function $E(n_B)$ around the minimum $n_0$ gives
\begin{equation}
    E(n_B)=E_0+\frac{1}{2} K_0 \frac{(n_B-n_0)^2}{9 n_0} + \mathcal{O}\left((n_B-n_0)^3 \right).
\end{equation}
As always the local minimum which is identified with the nuclear saturation point with saturation energy $E_0$.
The curvature of the energy curve is controlled by the compression modulus $K_0$ and determines the increase in energy due to compression.
For our choice of the coupling constants~\eqref{eq: Coupling constants} the saturation energy per baryon and saturation density are respectively $E_0=912$ MeV and $n_0=0.160$ fm$^{-3}$,  which almost perfectly corresponds to the physical values of the saturation energy and density.
An important observation is that the difference between the energy at nuclear saturation and the classical energy at zero density is much smaller than in previous works.
Indeed, the difference is now roughly $\Delta{E}\approx 7$ MeV, which is a $0.8\%$ difference with respect to the total energy.
Whereas, for a $B=32$ or $B=108$ $\alpha$-crystal the difference is found to be approximately $3\%$ and $1.7\%$, respectively.
This small difference in energy between the nuclear saturation and low-density asymptotic solutions is crucial for the existence of a purely skyrmion generated EoS at all densities.

Unfortunately, the compression modulus still exceeds the experimental value by a factor of $6\sim 7$. 
Although, in comparison with studies involving the $\textup{SC}_{1/2}$ crystal, where $K_0 \sim 1700$ MeV, we do observe a significant improvement in the (in)compressibility by approximately $500$ MeV ($K_0=1169)$, the non-homogeneous solution alone cannot solve the compression modulus problem in the Skyrme model.  
Nevertheless, this negative result is very important as it shows that the purely pionic Skyrme model cannot lead to a physically acceptable compression modulus.
Consequently, it seems to be necessary to include other mesonic d.o.f. which may soften the skyrmionic matter at the saturation point.


\section{Quantum skyrmion crystals and the symmetry energy}
\label{sec: Quantum of Skyrme crystals and the symmetry energy}

In general, the full symmetry group of the generalized $\mathcal{L}_{0246}$-Lagrangian \eqref{eq: Generalized Lagrangian} is the direct product of the Poincar\'e group and chiral group: $\tilde{G} = \textup{O}(3) \ltimes \mathbb{R}^3 \times \SO(4)_\textup{chiral}$.
However, static energy minimizers break the Poincar\'e symmetry group $\textup{O}(3) \ltimes \mathbb{R}^3$ to the Euclidean subgroup $E_3 = \SO(3) \times \mathbb{R}^3$, corresponding to spatial translations and rotations.
The resulting symmetry group of the static energy functional \eqref{eq: Skyrme model - Static energy} is thus $G = E_3 \times \SO(4)_\textup{chiral} \cong E_3 \times \SU(2)_L \times \SU(2)_R$.
The action of this group on the Skyrme field is given by
\begin{equation}
\label{eq: General action}
    \varphi(x) \mapsto A_L \varphi(Rx + X) A_R^\dagger,
\end{equation}
where $A_{L/R} \in \SU(2)_{L/R}$, $R \in \SO(3)$ and $X \in \mathbb{R}^3$.

For skyrmions on $M=\mathbb{R}^3$, one must impose finite boundary conditions $\varphi(x \rightarrow \infty) = \mathbb{I}_2$.
This allows for the compactification of the domain $\mathbb{R}^3 \bigcup \{\infty\} \cong S^3$ and further reduces the symmetry group $G$ to the subgroup $H = E_3 \times \diag[\SU(2)_L \times \SU(2)_R] \cong E_3 \times \SU(2)_I$, where $\SU(2)_I$ is the isospin internal symmetry group.
The corresponding action of the subgroup $H$ on the Skyrme field is given by the transformation \eqref{eq: General action} with $A_L=A_R=A \in \SU(2)_I$.

When considering crystals on $M=\mathbb{R}^3/\Lambda$, one must be careful when defining the isospin subgroup $\SU(2)_I$; the vacuum boundary condition is no longer imposed and there is not a natural way to select the diagonal isospin subgroup $\SU(2)_I$.
This problem was addressed by Baskerville \cite{Baskerville_1996} in the context of the $\textup{SC}_{1/2}$ crystal in the $\mathcal{L}_{24}$-model, wherein she considered full $\SO(4)_\textup{chiral}$ rotations.
She deduced that there are two cubic point groups that can define the $\textup{SC}_{1/2}$ crystal, one of which is related to the centers of the half-skyrmions.
The cubic point group symmetry corresponding to the half-skyrmion centers is reducible into the trivial $1$-dimensional irreducible representation and a $3$-dimensional irrep.
We choose the $\sigma=\varphi^0$ field to transform in the $1$-dimensional irrep.
Then the isospin group $\SU(2)_I$ can be defined as the group of isorotations of the pion fields $\vec{\pi}=(\varphi^1,\varphi^2,\varphi^3)$, corresponding to transformations in the $3$-dimensional irrep.
If the pion mass potential term $\mathcal{L}_0$ is included then this is a natural choice of isospin group $\SU(2)_I$.


\subsection{Isospin quantization}
\label{subsec: Isospin quantization}

As a field theory, the Skyrme model is non-renormalizable.
One must quantize it semi-classically.
It is well-known that the classical dynamics of slowly moving solitons corresponds to geodesic motion on the moduli space of static configurations \cite{Manton_1982}.
Minimal energy configurations in the Skyrme model are unique, for a given baryon number $B$, up to actions of the symmetry group $H=E_3 \times \SU(2)_I$.
The classical configuration space $Q$ of skyrmions is split into connected components, labeled by the baryon number $B$, $Q=\bigcup_{B\in\mathbb{Z}}Q_B$.
The covering space $\tilde{Q}_B$ of each component is a double-cover with a two-to-one map $\pi_Q: \tilde{Q}_B \rightarrow Q_B$ \cite{Krusch_2003}.
It was argued by Finkelstein and Rubinstein \cite{Finkelstein_1968} that the wave functions $\Psi \in \mathcal{H}$ must be defined on the covering space of the configuration space $\tilde{Q}$, where $\mathcal{H}$ is a formal Hilbert space such that $\Psi$ is normalizable and square integrable.
That is, the wave functions are defined by the map $\Psi: \tilde{Q} \rightarrow \mathbb{C}$.
We make a simple approximation and require the wave function $\Psi$ to be non-vanishing only on minimal energy configurations and their symmetry orbits.
This quantization procedure is known as rigid-body, or zero mode, quantization.

In the zero mode quantization method, a skyrmion is treated as a rigid body that is free to translate and rotate in physical space and also rotate in isospace, with the action defined by \eqref{eq: General action}.
These solutions are all degenerate in energy and this classical degeneracy is removed when one quantizes the theory.
We wish to quantize the isorotational degrees of freedom and work in the zero-momentum frame, ignoring the translational and spin degrees of freedom.
The action of the group of isorotations $\SU(2)_I$ on the Skyrme field $\varphi$ is defined by the mapping $\varphi(x) \mapsto A \varphi(x) A^\dagger$.
Semi-classical quantization is performed by promoting the the collective coordinate $A\in\SU(2)$ to a dynamical degree of freedom $A(t)$ \cite{Braaten_1988}.
The dynamical ansatz for the Skyrme field is then given by the transformation
\begin{equation}
    \varphi(x) \mapsto \hat{\varphi}(x,t) = A(t) \varphi(x) A^\dagger(t).
\label{eq: RBQ - Dynamical ansatz}
\end{equation}
Define the isorotational angular velocity $\vec{\omega}$ to be $A^\dagger \dot{A} = \frac{i}{2} \omega_j \tau^j$ such that $\omega_j = -i \Tr(\tau^j A^\dagger \dot{A})$.
Then, under the dynamical ansatz~\eqref{eq: RBQ - Dynamical ansatz}, the Maurer-Cartan left current transforms as
\begin{equation}
    \hat{L}_\mu = \hat{\varphi}^\dagger \partial_\mu \hat{\varphi} =
    \begin{cases}
    A \omega_i T_i A^\dagger, & \mu=0\\
    A L_i A^\dagger, & \mu=i=1,2,3,
    \end{cases}
\end{equation}
where $T_i = \frac{i}{2} \varphi^\dagger [\tau^i,\varphi]$ is an $\su(2)$-valued current.

The dynamical ansatz~\eqref{eq: RBQ - Dynamical ansatz} induces a rotational kinetic term in the energy functional, which is given by
\begin{align}
    E_\textup{rot} = \, & \int_{\mathbb{T}^3} \left\{ -\frac{c_2}{2} \Tr\left(\hat{L}_0\hat{L}_0\right) - \frac{c_4}{2} g^{ij} \Tr\left([\hat{L}_0,\hat{L}_i] [\hat{L}_0,\hat{L}_j]\right) \right. \nonumber \\
    \, & \left. + \frac{c_6}{g} g_{ij} \hat{\mathcal{B}}^i \hat{\mathcal{B}}^j \right\} \textup{vol}_g,
\end{align}
where the Chern--Simons current transforms as
\begin{equation}
    \hat{\mathcal{B}}^i = \frac{3}{24\pi^2} \epsilon^{ijk} \Tr(\hat{L}_0\hat{L}_j\hat{L}_k) = \frac{1}{8\pi^2} \epsilon^{ijk} \Tr(T_l L_j L_k) \omega_l.
\end{equation}
The restriction of the kinetic energy functional of the model to the isospin orbit of a given static solution defines a left invariant metric on $\SO(3)$ called the isospin inertia tensor, which is the symmetric $3 \times 3$-matrix given by
\begin{equation}
\label{eq: Isospin inertia tensor}
    U_{ij} = \int_{\mathbb{T}^3} \textup{d}^3x \sqrt{-g} \, \mathcal{U}_{ij},
\end{equation}
where the isospin inertia tensor density is
\begin{equation}
\label{eq: Isospin inertia tensor density}
\begin{split}
    \mathcal{U}_{ij} = -c_2 \Tr\left(T_i T_j\right) - c_4 g^{kl} \Tr\left([T_i,L_k][T_j,L_l]\right) \\ + c_6 g^{kl} \frac{\epsilon^{kab}\epsilon^{lcd}}{(4 \sqrt{2} \pi^2 \sqrt{-g})^2} \Tr(T_i L_a L_b) \Tr(T_j L_c L_d).
\end{split}
\end{equation}
Therefore, the effective Lagrangian on this restricted space of configurations is $L_{\textup{eff}} = L_{\textup{rot}} - M_B$, where $M_B$ is the static mass of the skyrmion defined by \eqref{eq: Skyrme model - Static energy} and $L_{\textup{rot}}$ is the induced isorotational part of the Lagrangian given by
\begin{equation}
    L_{\textup{rot}} = \frac{1}{2} \omega_i U_{ij} \omega_j.
\end{equation}

The rigid-body wavefunctions will be on $\SU(2)$ with isospin half-integer if $\mathcal{B}$ is odd and integer if $\mathcal{B}$ is even.
The isorotation angular momentum operator canonically conjugate to $\vec{\omega}$ is the body-fixed isospin angular momentum operator $\vec{K}$, defined by
\begin{equation}
    K_i = \partial L_{\textup{rot}}/ \partial \omega_i = U_{ij} \omega_j.
\end{equation}
This is related to the usual space-fixed isospin angular momentum $\vec{I}$ by the relation
\begin{equation}
    I_i = -D(A)_{ij} K_,
\end{equation}
where $A\in\SU(2)$ has been recast in the $\SO(3)$ form via the map
\begin{equation}
    D: \SU(2) \rightarrow \SO(3), \quad D(A)_{ij} = \frac{1}{2}\Tr\left(\tau^i A \tau^j A^\dagger \right).
\end{equation}
These two classical momenta are promoted to quantum operators $\vec{\hat{K}}$ and $\vec{\hat{I}}$, both satisfying the $\su(2)$ commutation relations, and the Casimir invariants satisfy $\vec{\hat{I}}^2=\vec{\hat{K}}^2$.
On the double cover of the group of isorotations $\SU(2)_I$, there is a basis of rigid-body wavefunctions $\ket{I,I_3,K_3}$ with $-I\leq K_3 \leq I$, where $I$ is the total isospin quantum number, $K_3$ is the third component of $\vec{\hat{K}}$ and $I_3$ is the third component of isospin relative to the space-fixed axes (in units of $\hbar$) as defined in nuclear physics.
The operator $\vec{\hat{I}}^2$ has eigenvalue $I(I+1)$ and $I_3$ the eigenvalue for the operator $\hat{I}_3$.
The rigid-body Hamiltonian takes the general form
\begin{equation}
    \mathscr{H} = \frac{\hbar^2}{2} \vec{\hat{K}} U^{-1} \vec{\hat{K}}^T + M_B.
\end{equation}
For Skyrme crystals, we can set the principal axes of inertia to be the usual orthogonal axes such that $U_{ij}=0$ for $i \neq j$.
Let us label the eigenvalues $U_i=U_{ii}$, then the quantum Hamiltonian takes the form
\begin{equation}
    \begin{split}
        \mathscr{H} = \frac{\hbar^2}{2}  \left( \frac{1}{U_1} + \frac{1}{U_2} \right) \vec{\hat{K}}^2 + \frac{\hbar^2}{2} \left( \frac{1}{U_3} - \frac{1}{U_1} - \frac{1}{U_2} \right) \hat{K}_3^2 \\ - \frac{\hbar^2}{2U_2} \hat{K}_1^2 - \frac{\hbar^2}{2U_1} \hat{K}_2^2 + M_B.
    \end{split}
\label{eq: Diagonal crystal Hamiltonian}
\end{equation}
The energy eigenstates of the Hamiltonian~\eqref{eq: Diagonal crystal Hamiltonian} can be classified by $I$ and $I_3$.
To determine bound states with definite energy one must solve the static Schr\"{o}dinger equation corresponding to this Hamiltonian, $\mathscr{H}\ket{\Psi} = E\ket{\Psi}$.
The Schr\"{o}dinger equation can be expressed more explicitly within a particular $(I,I_3)$ sector by expanding the quantum state $\ket{\Psi}$ in terms of the total wavefunctions $\Psi$ as
\begin{equation}
    \ket{\Psi} = \sum_{K_3=-I}^{+I} \Psi_{K_3}(q) \ket{I,I_3,K_3}, \quad \vec{\Psi}(q) = \begin{pmatrix}
        \Psi_{-I}(q) \\
        \vdots \\
        \Psi_{+I}(q)
    \end{pmatrix},
\end{equation}
with $q \in \tilde{Q}$ and substituting this into the Hamiltonian~\eqref{eq: Diagonal crystal Hamiltonian}.


\subsection{Symmetry energy and the cusp structure}
\label{subsec: Symmetry energy and the cusp structure}

So far we have only considered symmetric nuclear matter, which we have described by using the classical multi-wall skyrmion crystal.
In order to study nuclear matter in neutron stars we must consider isospin asymmetric nuclear matter, whereby a small fraction of protons are permitted.
Now let us consider asymmetric nuclear matter with baryon number $B=N+Z$, where $N$ is the number of neutrons and $Z$ the number of protons.
The asymmetry of such matter is determined by the isospin asymmetry parameter $\delta = (N-Z)/(N+Z)=1-2\gamma$, where $\gamma$ is the proton fraction.
We define the nuclear density to be $n_B = B/V$, with the nuclear saturation density $n_0$ defined to be the nuclear density such that $(\partial M_B)/(\partial n_B)|_{n_B=n_0}=0$.
Then the binding energy per baryon number of asymmetric nuclear matter is given by
\begin{equation}
    \frac{E}{B}(n_B,\delta) = E_N(n_B) + S_N(n_B) \delta^2 + \textup{O}(\delta^3).
\label{eq: Asymmetric binding energy}
\end{equation}

The two terms appearing in the asymmetric binding energy~\eqref{eq: Asymmetric binding energy} are the binding energy of isospin-symmetric matter $E_N$ and the symmetry energy $S_N$.
In terms of our model, the symmetric binding energy is defined by $E_N = (M_B - BM_1)/B$.
The symmetry energy $S_N$ dictates how the binding energy changes when going from symmetric $(\delta=0)$ to asymmetric $(\delta \neq 0)$ nuclear matter.
We can expand the isospin symmetric binding energy $E_N$ and the symmetry energy $S_N$ around the saturation density $n_0$ for symmetric matter \cite{Fantina_2018},
\begin{align}
    E_N(n_B) = \, & E(n_0) + \frac{1}{18} K_0 \epsilon^2, \\
    S_N(n_B) = \, & S_0 + \frac{1}{3}L_\textup{sym} \epsilon + \frac{1}{18} K_\textup{sym} \epsilon^2 + \textup{O}(\epsilon^3),
\end{align}
where $\epsilon=(n_B-n_0)/n_0$, $K_0$ is the incompressibility at the saturation point and $S_0=S_N(n_0)$ is the symmetry energy coefficient at saturation.
We remind ourselves that, for our choice of coupling constants~\eqref{eq: Coupling constants}, the nuclear saturation point is characterized by the density $n_0=0.160\,\textup{fm}^{-3}$ and energy (per baryon) $M_B/B=912$ MeV.  
The higher-order coefficients, $L_\textup{sym}$ and $K_\textup{sym}$, appearing in the symmetry energy $S_N$ are defined as
\begin{equation}
    L_\textup{sym} = 3 n_0 \left. \frac{\partial S_N}{\partial n_B} \right|_{n_B=n_0}, \quad 
    K_\textup{sym} = 9 n_0^2 \left. \frac{\partial^2 S_N}{\partial n_B^2} \right|_{n_B=n_0}.
\end{equation}
The precise values of these coefficients are not known, but are predicted to be $L_\textup{sym} = 57.7 \pm 19$ MeV and $K_\textup{sym} = -107 \pm 88$ MeV \cite{Zhang_2021}.

Consider an infinitely extended and rigidly iso-spinning Skyrme crystal with each unit cell containing baryon number $B_{\textup{cell}}$.
In order to calculate the isospin correction to the energy of the crystal we would need to know the quantum state of the whole crystal.
This is obviously a very difficult computation since the crystal is infinitely extended and is therefore composed of an infinite number of baryons.
However we may impose the following restrictions to solve this problem:
\begin{itemize}
    \item The total isospin quantum state of the crystal $\ket{\Psi}$ is written as the superposition of each individual unit cell state $\ket{\psi}$.
    That is $\ket{\Psi} = \otimes_{N_\textup{cell}} \ket{\psi}$, where $N_\textup{cell} \rightarrow \infty$ in the thermodynamic limit.
    \item The symmetry of the classical configuration in each unit cell is extended to the whole crystal, so both wavefunctions share the same point symmetry group.
\end{itemize}
Under these assumptions, and since we have $B_\textup{cell}=4$ within our unit cell, there are a finite number of possible quantum states with allowed quantum numbers $I = 0, 1, 2$ \cite{Wereszczynski_2022}.
The $I_3 = 0$ case, which corresponds to symmetric nuclear matter, would be the one with the lowest energy since it has no isospin energy compared to the other cases.
This is obviously the most symmetric state possible.
However, it is known that inside neutron stars there is a huge asymmetry between protons and neutrons.
Baskerville investigated the charge neutral case $I_3 = -2$, corresponding to a pure neutron crystal, and computed the quantum isospin corrections to the energy \cite{Baskerville_1996}.
However, a realistic description of neutron stars would require the presence of protons.
Although the concrete value is still unknown, simulations yield values around $\gamma \sim 10^{-2}-10^{-1}$ \cite{PhysRevC.85.015807,PhysRevC.72.015802}.
Therefore, following the arguments in \cite{Wereszczynski_2022} we perform a mean-field approximation considering a larger chunk of crystal, enclosing an arbitrary number of unit cells $N_\textup{cell}$, which is in a generic quantum state with fixed eigenvalue,
\begin{equation}
    I_3 = \frac{(Z - N)}{2} = -\frac{(1 - 2\gamma)}{2}N_{\textup{cell}}B_{\textup{cell}}.
    \label{MF_I3val}
\end{equation}
Note that in this case the nuclear density of the crystal chunk can be directly interpreted as the nuclear density of the unit cell, since
\begin{equation}
    n_B = \frac{B_\textup{crystal}}{V_\textup{crystal}} = \frac{N_{\textup{cell}} B_{\textup{cell}}}{N_\textup{cell} 
 V_{\textup{cell}}} = \frac{B_{\textup{cell}}}{
 V_{\textup{cell}}}.
\end{equation}

In previous applications of skyrmion crystals to model neutron stars (see, for example, \cite{sym15040899,Wereszczynski_2022,Wereszczynski_2023,Park_2010,Rho_2013}), the $\textup{SC}_{1/2}$ crystal was considered.
This crystal has an isotropic inertia tensor with eigenvalue $U_i = \lambda$, with $\lambda$ some constant.
However, the multi-wall crystal considered in this paper is not isotropic and the isospin inertia tensor generically has the eigenvalues $U_1 = U_2 \neq U_3$.
The Schr\"{o}dinger equation corresponding to such a rigidly iso-spinning crystal with $N_\textup{cell}$ unit cells can be written as
\begin{equation}
    \mathscr{H} \ket{\Psi} = \left(N_{\textup{cell}}M_B + E_{I,I_3}\right) \ket{\Psi},
\end{equation}
where the isospin correction to the energy of the crystal is given by
\begin{equation}
    E_{I,I_3} = \frac{\hbar^2I(I+1)}{N_{\textup{cell}}U_1} + \frac{\hbar^2I_3^2}{2} \left( \frac{1}{U_3} - \frac{2}{U_1} \right).
\end{equation}
It should be noted that in addition to the quantum numbers $I,I_3$ being density $n_B$ dependent, the inertia tensor is also density dependent, that is $U_i=U_i(n_B)$.

The eigenvalue $I_3$ is already fixed from the mean-field approximation \eqref{MF_I3val}, and the value of $I = I_3$ is the one which minimizes the isospin energy, since by definition $I^2 \geq I^2_3$.
In the thermodynamic limit $N_{\textup{cell}} \rightarrow \infty$ we obtain a final expression for the quantum correction (per unit cell) to the energy due to the isospin degrees of freedom,
\begin{equation}
    E_\textup{iso}(n_B) = \frac{\hbar^2}{8U_3(n_B)} B_{\textup{cell}}^2 \delta^2.
\label{eq: Quantum isospin energy}
\end{equation}
This quantum isospin energy is explicitly related to the proton fraction $\gamma$, and so we will need to include leptons if we are to allow the crystal to have a non-zero proton fraction.
This is required in order for the system to remain electrically neutral.
Thus the proton fraction, and hence the quantum state of the crystal, will be obtained by imposing $\beta$-equilibrium for each value of the density.

\begin{figure}[t]
    \centering
    \includegraphics[width=0.5\textwidth]{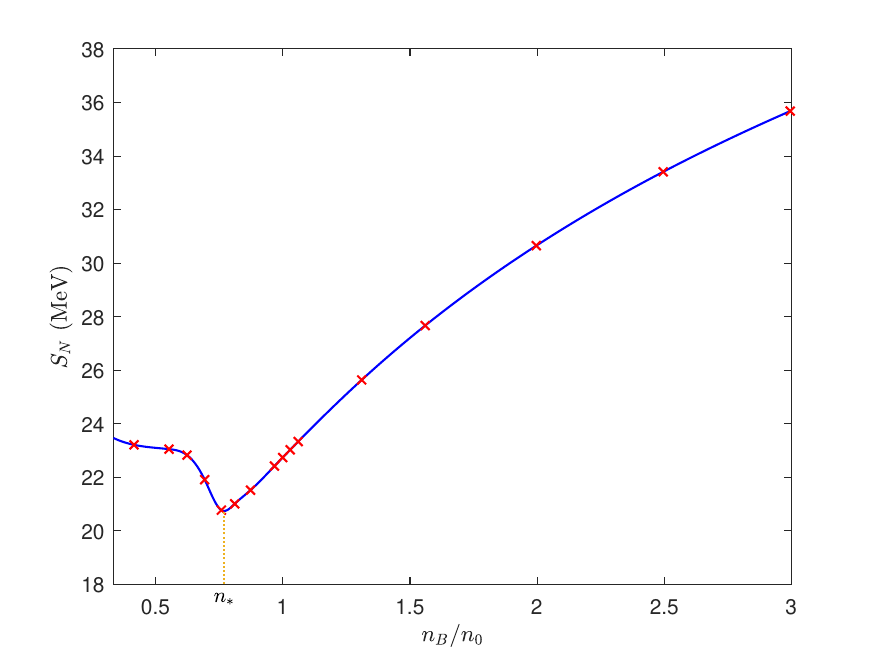}
    \caption{The nuclear symmetry energy $S_N$ as a function of the baryon density $n_B$, exhibiting the cusp structure detailed in the text at $n_*\sim 3n_0/4$.}
\label{fig: Plot_SymmetryEnergy}
\end{figure}

From the quantum isospin energy~\eqref{eq: Quantum isospin energy}, we can determine the nuclear symmetry energy of the multi-wall crystal, which in general plays a crucial role in the structure of neutron-rich nuclei and, of more interest to us, in neutron stars.
For general skyrmion crystals the symmetry energy is given by
\begin{equation}
    S_N(n_B) = \frac{\hbar^2}{8U_3(n_B)} V_{\textup{cell}} n_B,
\end{equation}
where the eigenvalue $U_3$ of the isospin inertia tensor~\eqref{eq: Isospin inertia tensor} is implicitly dependent on the nuclear density $n_B$.
We determine the symmetry energy at at saturation to be $S_0=22.7$ MeV, which is in okay agreement with the experimentally observed value $S_0\sim 30$ MeV \cite{FiorellaBurgio:2018dga}.
The resulting symmetry energy curve $S_N(n_B)$ for the multi-wall crystal is plotted in Fig.~\ref{fig: Plot_SymmetryEnergy}.
Having obtained the symmetry energy curve we can determine its slope and curvature, which are computed at the nuclear saturation point.
We find that they are, respectively, $L_\textup{sym}=36.6$ MeV and $K_\textup{sym}=-15.1$ MeV.

Let us now summarize the results obtained for the multi-wall crystal.
First of all, we find that at lower densities the isospin moment of inertia, and specifically its eigenvalue $U_3$, tends to a constant value.
This is an obvious consequence of the inhomogeneous nature of the solution which, in the limit $V_\textup{cell} \rightarrow \infty$, tends to an ``isolated'' multi-wall configuration on $M=S^1 \times S^1 \times \mathbb{R}$.
This simple fact has an important consequence.
Namely, it leads to a non-zero value of the symmetry energy at zero density, $S_N(0)=23.8$ MeV.
At a first glance, this seems to be in contradiction with the standard description of nuclear matter where the symmetry energy vanishes at zero density.
However, we want to argue that this is a desirable property of the Skyrme model as it indicates a smooth transition between infinite nuclear matter and finite atomic nuclei.
Indeed, the asymmetry energy in the Bethe–Weizs\"{a}cker SEMF reads 
\begin{equation}
    E_\textup{asym} = a_{A}\frac{(N-Z)^2}{B} = a_A \delta^2 B,
\end{equation}
where $a_A\approx 23.7$ MeV.
Thus, our symmetry energy at zero density can be directly identified with $a_A$ with excellent agreement. 

We remark that the assumed identification here between the zero density symmetry energy and the asymmetry energy in the Bethe-Weizs\"{a}cker formula is not a unique possibility.
In fact, in the seminal paper by Natowitz el. al. \cite{Natowitz:2010ti} they computed the symmetry energy of the low density, warm nuclear matter using a quantum-statistical approach.
Their results agree amazingly well with values extracted from heavy-ion collisions \cite{Kowalski:2006ju}.
The symmetry energy, still taking a non-zero value at zero density, is approximately only one fourth of its value at saturation $n_0$.
It would definitely be very desirable to investigate whether the Skyrme model may lead to similar results or not.

Moving away from zero nuclear density towards $n_*\sim 3n_0/4$, the isospin energy and consequently the symmetry energy slowly decreases, as can be seen in Fig. \ref{fig: Plot_SymmetryEnergy}.
This again is not an unexpected result in the Skyrme model.
It was noticed by Kopeliovich \textit{et al.} \cite{Kopeliovich:2004pd} that the careful analysis of mass splittings of nuclear isotopes leads to the symmetry energy decreasing with increasing baryon number $B$.
Here, we reproduce this result, however, using a completely different setup, i.e. the collective coordinate quantization of the crystal ground state. 

Below the nuclear saturation point $n_0$ at the density $n_*\sim 3n_0/4$, the symmetry energy exhibits a \textit{cusp} structure.
This cusp also seems to be a generic feature of the Skyrme model, independent of the choice of values for the coupling parameters~\eqref{eq: Coupling constants} but rather can be interpreted as the point where the multi-wall crystal begins transitioning to an ``isolated'' multi-wall.
On the other hand, its position with respect to the saturation point certainly may be affected by a choice of the model parameters.
One can also expect such a cusp to be present where a crystalline configuration transitions to an isolated configuration at zero nuclear density, e.g. for the $\alpha$ and chain crystals.
It is interesting to remark that such a cusp, albeit above the saturation density $n_B > n_0$, has been advocated in \cite{Rho_2011, Lee:2021hrw} as an effect of an assumed topological phase transition from the FCC crystal of $B=1$ hedgehogs to the $\textup{SC}_{1/2}$ crystal of fractional skyrmions as the nuclear density grows.
Although, in reality such a transition does not occur in the Skyrme model as it is found to occur in the thermodynamically unstable regime $n_B < n_0$ \cite{Adam_2022}.
To conclude our findings on the symmetry energy cusp, we propose that the origin of the cusp can be associated with a phase transition between an \textit{infinite crystalline} state and a somewhat \textit{isolated} state that is \textit{non-homogeneous} and \textit{nucleated}.


\section{Particle fractions of $npe\mu$ matter in $\beta$-equilibrium}

For a more realistic description of cold nuclear matter inside neutron stars we need to consider not completely asymmetric nuclear matter.
As was shown in the previous section, this can be achieved by allowing a small fraction of protons over neutrons.
The presence of protons gives the crystal positive electric charge, so we need to include a background of negatively charged leptons to neutralize the system.
To determine the proton fraction $\gamma$ at a prescribed nuclear density $n_B$ we impose charge neutrality and $\beta$-equilibrium conditions, and then we solve the underlying equilibrium equation.
Additionally, the presence of protons would require the inclusion of Coulomb interaction within the unit cell and between neighbouring cells.
It has been argued \cite{Klebanov_1985} that the contribution of this energy diverges in the crystal due to infinitely many interactions between the cells.
However, including a background of negatively charged particles in the system suppresses the Coulomb interaction between neighbouring cells and hence has a negligible contribution to the energy \cite{Wereszczynski_2022}.

In the neutron star interior, the interaction between leptons and nuclear matter is mediated by the weak force.
We can describe the exchange of leptons and nucleons by electron capture and $\beta$-decay processes, respectively,
\begin{subequations}
    \begin{align}
        p + l &\rightarrow n + \nu_l \\
        n &\rightarrow p + l + \bar{\nu}_l.
    \end{align}
\end{subequations}
These processes take place simultaneously at the same rate, assuming that the charge neutrality,
\begin{equation}
    n_p = \frac{Z}{V} = n_e + n_\mu,
\label{eq: Charge neutrality}
\end{equation}
and the $\beta$-equilibrium conditions \cite{Glendenning_1997},
\begin{equation}
    \mu_p = \mu_n - \mu_I \quad \Rightarrow \quad \mu_I = \mu_l, \quad l=e,\mu,
\label{eq: Beta-equilibrium}
\end{equation}
are satisfied.
Here $\mu_I$ is the isospin chemical potential given by
\begin{equation}
    \mu_I = \frac{\delta B\hbar^2}{2U_3} = \frac{(1-2\gamma)B\hbar^2}{2U_3}.
\end{equation}
Leptons inside a neutron star are treated as a non-interacting, relativistic, highly degenerate Fermi gas.
The corresponding chemical potential for such a type of lepton is given by \cite{Wereszczynski_2023}
\begin{equation}
    \mu_l = \sqrt{(\hbar k_F)^2 + m_l^2},
\end{equation}
where $k_F=(3\pi^2n_l)^{1/3}$ is the associated Fermi momentum and $m_l$ the lepton mass.
For electrons we take the ultra-relativistic approximation $\mu_e \approx \hbar k_{F,e}$.
From the charge neutrality condition \eqref{eq: Charge neutrality}, the electron number density is
\begin{equation}
    n_e = \frac{\gamma B}{V} - n_\mu.
\end{equation}
The $\beta$-equilibrium condition \eqref{eq: Beta-equilibrium} for electrons yields the following relation
\begin{equation}
    \mu_I = \mu_e \quad \Rightarrow \quad \frac{\hbar B (1-2\gamma)}{2U_3} = \left[ 3\pi^2 \left( \frac{\gamma B}{V} - n_\mu \right) \right]^{1/3},
\label{eq; Electron chemical potential relation}
\end{equation}
and for muons gives
\begin{equation}
    \mu_I = \mu_\mu \quad \Rightarrow \quad n_\mu = \frac{1}{3\pi^2}\left[ \left( \frac{\hbar B (1-2\gamma)}{2U_3} \right)^2 - \left( \frac{m_\mu}{\hbar} \right)^2 \right]^{3/2}.
\label{eq; Muon chemical potential relation}
\end{equation}
In the low density regime the electron chemical potential will be smaller than the muon mass, $\mu_e < m_\mu$.
So we can solve \eqref{eq; Electron chemical potential relation} in the low density regime considering only electrons, by setting $n_\mu=0$ until $\mu_e \geq m_\mu$.
Once the electron chemical potential $\mu_e$ reaches the muon mass $m_\mu = 105.66$ MeV at high densities, it will be energetically favourable for muons to appear.
Then we solve \eqref{eq; Electron chemical potential relation} and \eqref{eq; Muon chemical potential relation} simultaneously \cite{Wereszczynski_2023}, and construct the proton fraction curve $\gamma = \gamma(n_B)$. 

\begin{figure}[t]
    \centering
    \includegraphics[width=0.5\textwidth]{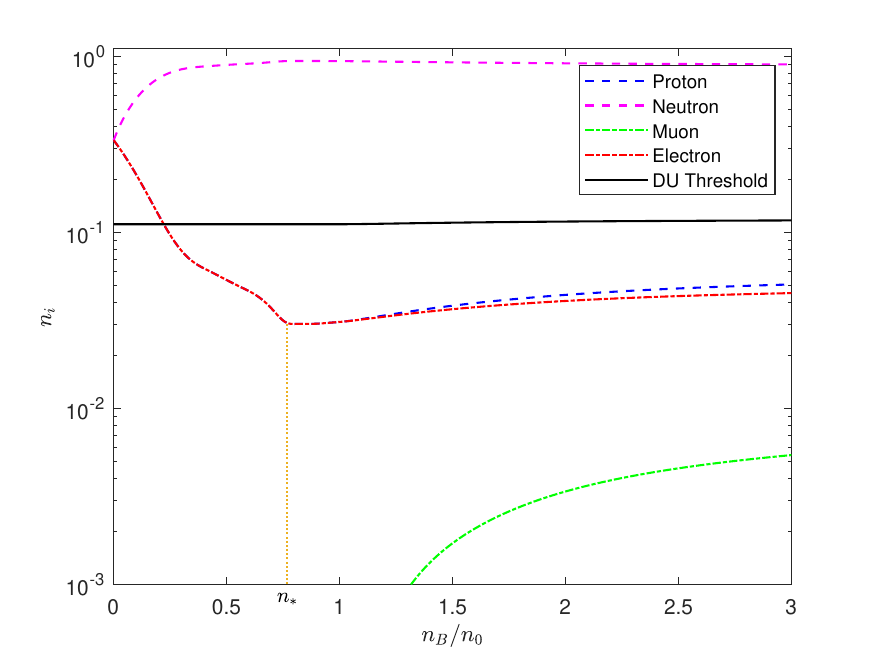}
    \caption{Plot of the particle number densities $n_i$ as functions of the baryon density $n_B$.
    The particle number densities are normalized such that the total number density is $\sum_{i}n_i = 1$.
    The transition between isospin asymmetric infinite matter and symmetric finite matter at the cusp density $n_*$ is now manifest.}
\label{fig: Plot_ParticleFractions}
\end{figure}

In Fig. \ref{fig: Plot_ParticleFractions} we plot the particle fractions of $npe\mu$ matter in $\beta$-equilibrium for the multi-wall crystal.
Note that the cusp structure present in the symmetry energy, or equivalently in the isospin energy, results in an appearance of a similar structure in the particle fractions.
This reinforces the proposition that the cusp density point $n_*$ is the density at which a phase transition between isospin asymmetric infinite nuclear matter and symmetric finite matter begins.
Furthermore, the fact that the symmetry energy $S_N$ tends to a constant value at zero density leads to a similar behavior for the proton, neutron and electron particle fractions.
Namely, they take their minimal/maximal value at $n_*$ then they increase/decrease as zero density is approached.
This is once again a direct consequence of a non-zero value of the isospin moment of inertia at this limit and, therefore, a generic feature of the Skyrme model.
We remark that at zero density $n_B=0$, which, in the Skyrme model framework, can be interpreted as a limit where we find nuclei in the vacuum, the nuclear matter becomes totally isospin symmetric with $\gamma_p(0)=0.5$.
This corresponds quite well to the proton fraction in $^{56}$Fe, $\gamma_p=0.46$, which is the element expected to be present in the crust of neutron stars \cite{Chamel:2008ca}.
Further, it appears that there is a phase transition at ($n/n_B=0.91$, $p=0.023 \textup{MeV fm}^-3$).
The $n_*$ density occurs in this region of constant pressure, so it could very well be related to the liquid-gas phase transition.

We remark that at the high density, which corresponds to the core of neutron star, the proton fraction is quite small.
This agrees with previous computations in the Skyrme model with the $\textup{SC}_{1/2}$ crystal \cite{Wereszczynski_2022}.
Fortunately, inclusion of strange d.o.f. resolves this issue and brings the proton fraction to the widely accepted $\sim 0.4$ value, see \cite{Wereszczynski_2023}.
We expect that the same mechanism applies for the multi-wall crystal.
Especially considering this ground state crystalline solution and the $\textup{SC}_{1/2}$ crystal are basically identical at high density.
On the other hand, inclusion of Kaon condensate does not have any impact on the low density regime.

We now summarize our findings and compute the total energy per unit cell in a $\beta$-equilibrated multi-wall skyrmion crystal, that is
\begin{equation}
    E_\textup{cell}(\gamma) = M_B(\gamma) + E_\textup{iso}(\gamma) + E_e(\gamma) + E_\mu(\gamma),
\end{equation}
where the isospin energy for a $\beta$-equilibrated crystal is given by
\begin{equation}
    E_\textup{iso}(\gamma) = \frac{\hbar^2B_{\textup{cell}}^2}{8U_3} (1-2\gamma)^2.
\end{equation}
The lepton energies are the energies of a relativistic Fermi gas at zero temperature,
\begin{align}
    E_l = \, & \frac{V}{\hbar^3\pi^2} \int_0^{\hbar k_F} k^2\sqrt{k^2 + m_l^2} \, \textup{d}k \nonumber \\
    = \, & \frac{Vm_l^4}{8\hbar^3\pi^2} \left[ \frac{\hbar k_F}{m_l} \left(1+2\left(\frac{\hbar k_F}{m_l}\right)^2\right)\sqrt{\left(\frac{\hbar k_F}{m_l}\right)^2+1} \right. \nonumber \\
    \, & \left. - \sinh^{-1}\left(\frac{\hbar k_F}{m_l}\right) \right].
\end{align}

\begin{figure}[t]
    \centering
    \includegraphics[width=0.5\textwidth]{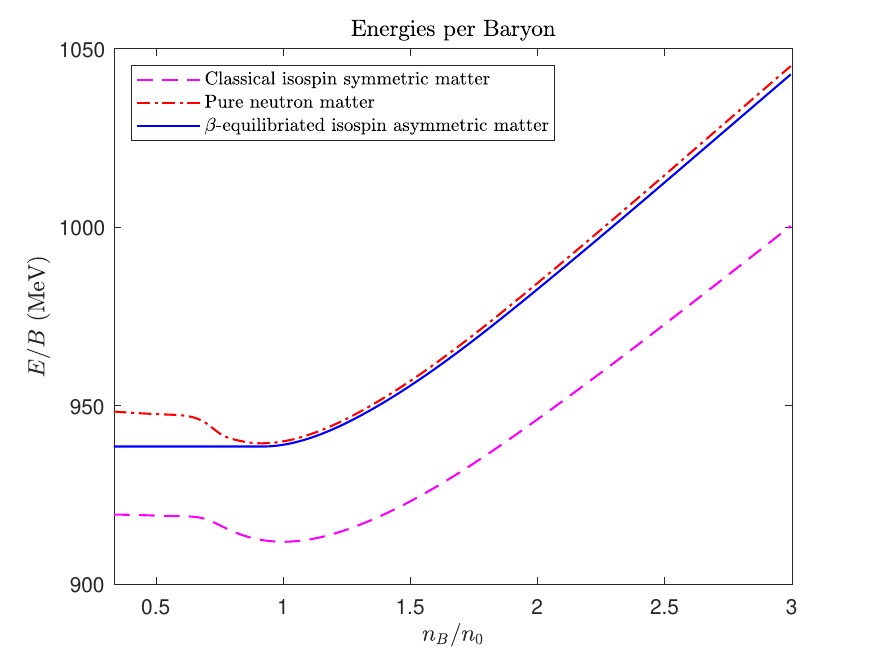}
    \caption{Comparison between the classical isospin symmetric crystal, the pure neutron crystal, and the $\beta$-equilibrated asymmetric crystal with the Maxwell construction applied.}
\label{fig: Plot_MaxwellConstruction}
\end{figure}

The crucial observation is that, in the case of the multi-wall skyrmion crystal, the inclusion of the $\beta$-equilibrated isospin energy and lepton energies does not completely erase the small minimum in the classical energy $M_B$.
Strictly speaking there is still a very shallow minimum at a density smaller than the saturation density, $n_B=0.146\,\textup{fm}^{-3}$.
For smaller densities the total energy grows, until a small maximum is reached. After that the total energy decreases as the nuclear density approaches the zero density limit, $n_B \rightarrow 0$.
Importantly, the asymptotic value of the total energy per unit cell is smaller than the energy at the minimum.
This means that, although the total energy per unit cell still possesses a thermodynamically unstable region, we can take advantage of the Maxwell construction and derive an EoS which is valid at all densities.
This is a valid construction and has a minute affect on the EoS since the difference in energy between the asymptotic solution and the minimum is $\Delta E \sim 0.1$ MeV.
The formulation of the Maxwell construction is detailed below and the resulting $\beta$-equilibrated asymmetric nuclear matter is plotted in Fig.~\ref{fig: Plot_MaxwellConstruction}, alongside the classical isospin symmetric matter and the pure neutron matter.

The pure neutron matter is obtained for the entirely isospin asymmetric case $\delta=1$ with $I_3=-2$.
Unlike the $\beta$-equilibrated matter, the pure neutron matter approaches the asymptotic solution from below.
This is due to the non-vanishing of the quantum isospin energy contributions $E_{\textup{iso}}(n_B)$ in the zero density limit $n_B \rightarrow 0$.
Consequently, the Maxwell construction cannot be used on the pure neutron matter EoS.

We remark that for the $\alpha$-crystal the total energy in the zero density limit is greater than the energy at the minimum, so the Maxwell construction is not possible.
On the other hand, for $B=32$ and $B=108$ crystals constructed from $\alpha$-particles, such a construction is possible but it extends over a non-physical range of densities and occurs for relatively high values of the pressure.
For example, the neutron stars obtained from these crystals would almost be entirely made from the Maxwell construction phase.

The Maxwell construction (MC), or equal area rule, is implemented as follows.
We find three points $V_1$, $V_2$ and $V_\textup{int}$ on the $E_\textup{cell}(V_\textup{cell})$ curve, with $V_1 < V_\textup{int} < V_2$, that have the same gradient/pressure, i.e. $p(V_i)=:p_{\textup{MC}}$.
These three points are chosen such that the area enclosed between $p([V_1,V_\textup{int}])$ and $p_{\textup{MC}}$ is equal to the area enclosed between $p([V_\textup{int},V_2])$ and $p_{\textup{MC}}$, where $p([V_1,V_\textup{int}]) \leq p_{\textup{MC}}$ and $p([V_\textup{int},V_2]) \geq p_{\textup{MC}}$.
This ensures that the total energy of the thermodynamic system remains the same while implementing this construction.
Then, in the corresponding MC density regime $V_1<V_\textup{cell}<V_2$, the total energy function is replaced by a straight line connecting $E(V_1)$ and $E(V_2)$.
The resulting total energy per unit cell function can be summarized as
\begin{equation}
    E^\textup{MC}(V)=\left\{
\begin{array}{lc}
    E(V) &  V\leq V_1\\
     E(V_1)-p_{\textup{MC}}(V-V_1) \;\;\;  & V_1\leq V \leq V_2 \\
     E(V) & V \geq V_2 
\end{array}.
    \right.
\end{equation}

Now we are in a position to determine the EoS for the multi-wall configuration.
The multi-wall crystal EoS for isospin asymmetric nuclear matter can be obtained by defining the energy density $\rho$ and pressure $p$ as, respectively,
\begin{align}
    \rho = \, & \frac{E}{V} = \frac{E_\textup{cell}}{V_\textup{cell}} = \frac{n_B}{B}E_\textup{cell}, \\
    p = \, & -\frac{\partial E}{\partial V} = - \frac{\partial E_\textup{cell}}{\partial V_\textup{cell}} = \frac{n_B^2}{B} \frac{\partial E_\textup{cell}}{\partial n_B}.
\end{align} 
This EoS $\rho=\rho(p)$, generated purely from the generalized multi-wall skyrmion crystal, is valid at all densities.
In our case, the pressure at which the Maxwell construction is applied is quite small, $p_\textup{MX}=0.023$ MeV fm$^{-3}$, which corresponds to an energy difference of $\approx 0.1$ MeV over a large density range ($0.91n_0$ to $0.36 n_0$).
The resulting EoS is shown in Fig.~\ref{fig: neutron star properties}, alongside the EoS without the Maxwell construction applied. 

Although the obtained equation of state covers the full range of densities one has to be aware that the multi-wall crystal does not describe the low density regime in its entirety.
As we have already mentioned, to get a more realistic description of the crust the electrostatic interaction should be included.
This can have an impact on the structure and symmetry of the skyrmions, which could potentially lead to the appearance of other non-homogeneous solutions with different baryon numbers per unit cell. 


\section{Neutron stars from quantum skyrmion crystals coupled to gravity}
\label{sec: Neutron stars from quantum skyrmion crystals coupled to gravity}

In order to describe neutrons stars within the Skyrme framework, we need to couple the generalized Skyrme model to gravity.
We do this by introducing the Einstein--Hilbert--Skyrme action \cite{Luckock_1986}
\begin{equation}
    S = \frac{1}{16 \pi G}\int_\Sigma \textup{d}^4x \sqrt{-g} R  + S_\textup{matter},
\end{equation}
where $G=1.3238094\times10^{-42}\,\textup{fm\,MeV}^{-1}$ is the gravitational constant and $R$ the Ricci scalar.
The matter part of the Einstein--Skyrme action, $S_\textup{matter}$, describes matter in the interior of the neutron star.
It is well known that the interior of a neutron star is well described as a perfect fluid of nearly free neutrons and a very degenerate gas of electrons.
We exploit this and use a perfect fluid model such that the energy-momentum tensor takes the form
\begin{equation}
    T_{\mu\nu} = -\frac{2}{\sqrt{-g}}\frac{\delta S_\textup{matter}}{\delta g^{\mu\nu}} = \left(\rho + p\right) u_\mu u_\nu + p g_{\mu\nu},
\end{equation}
where the energy density $\rho$ and the pressure $p$ are related by the multi-wall crystal EoS $\rho=\rho(p)$. 


\subsection{The Tolman--Oppenheimer--Volkoff system}
\label{subsec: The Tolman--Oppenheimer--Volkoff system}

Our aim is to calculate the maximum permitted mass and radius for a neutron star described by our system, and obtain the mass-radius curve.
Therefore we have to solve the resulting Einstein equations for some particular choice of metric ansatz.
The simplest case is that of a static non-rotating neutron star.
We use a spherically symmetric ansatz of the spacetime metric, which in Schwarzschild coordinates reads \cite{Naya_2015}
\begin{equation}
    \textup{d}s^2 = -A(r)\textup{d}t^2 + B(r)\textup{d}r^2 + r^2\left(\textup{d}\theta^2 + \sin^2\theta \textup{d}\phi^2 \right).
\label{eq: Spherical metric ansatz}
\end{equation}
The mass and radius of the neutron star can be calculated by inserting this spherical metric ansatz into the Einstein equations
\begin{equation}
    G_{\mu\nu} = 8\pi G T_{\mu\nu},
\label{eq: Einstein equation}
\end{equation}
where $G_{\mu\nu}=R_{\mu\nu}-\frac{1}{2}Rg_{\mu\nu}$ is the Einstein tensor, and solving the resulting Tolman--Oppenheimer--Volkoff (TOV) equations,
\begin{subequations}
    \begin{align}
        \label{eq: A(r)}
        \frac{\textup{d}A}{\textup{d}r} = \, & A(r) r \left( 8\pi G B(r) p(r) - \frac{1-B(r)}{r^2}   \right), \\
        \label{eq: B(r)}
        \frac{\textup{d}B}{\textup{d}r} = \, & B(r) r \left( 8\pi G B(r) \rho(p(r)) + \frac{1-B(r)}{r^2}   \right), \\
        \label{eq: p(r)}
        \frac{\textup{d}p}{\textup{d}r} = \, & -\frac{p(r)+\rho(p(r))}{2A(r)}\frac{\textup{d}A}{\textup{d}r}.
    \end{align}
\label{eq: TOV system}
\end{subequations}

The resulting  TOV system involves 3 differential equations for $A$, $B$ and $p$, which must be solved for a given value of the pressure in the center of the neutron star ($p(0) = p_0$) until the condition $p(R_\textup{NS})=0$ is achieved.
The radial point $R_\textup{NS}$ at which the pressure vanishes defines the radius of the neutron star, and the mass $M$ is obtained from the Schwarzschild metric definition outside the star,
\begin{equation}
    B(R_\textup{NS}) = \frac{1}{1 - \frac{2MG}{R_\textup{NS}} }.
\label{eq: Schwarzchild mass}
\end{equation}
In order for the metric function $B(r)$ to be non-singular at $r=R_\textup{NS}$, the pressure $p(r)$ must obey $p'(R_\textup{NS})=0$.

The TOV system~\eqref{eq: TOV system} is solved via a central shooting method from some initial central pressure $p_0$ at $r=0$ until the edge of the star has been reached (corresponding to $p(R_\textup{NS})=0$).
The amount of matter contained at $r=0$ should be zero, which gives the boundary conditions $B(0)=A(0)=1$.
That is, the spacetime metric should approach the Minkowski metric towards the neutron star core.
We can simultaneously apply a 4\textsuperscript{th} order Runge--Kutta method to the system of IVPs \eqref{eq: B(r)}, \eqref{eq: p(r)}, for the initial conditions $B(0)=1$ and $p(0)=p_0$, until the condition $p(R_\textup{NS})=0$ is achieved.
This yields the metric function $B(r)$ and the pressure profile $p(r)$ satisfying the necessary boundary conditions.
Then the metric function $A(r)$ can be easily obtained by numerically integrating \eqref{eq: A(r)}.
The corresponding radius $R$ and the stellar mass $M=M(R_\textup{NS})$ can be extracted from the Schwarzschild definition \eqref{eq: Schwarzchild mass}.
Increasing the central pressure $p_0$ in succession corresponds to determining a sequence of neutron stars of increasing mass, until the mass limit is reached \cite{Glendenning_1997}.
The observational mass limit is approximately $2.5 M_\odot$ \cite{LIGOScientific:2020aai}, where the solar mass is $M_\odot = 1.116 \times 10^{60}\,\textup{MeV}$.


\subsection{Neutron star properties and the mass-radius curve}
\label{subsec: Neutron star properties and the mass-radius curve}

\begin{figure}[t]
    \centering
    \includegraphics[width=0.5\textwidth]{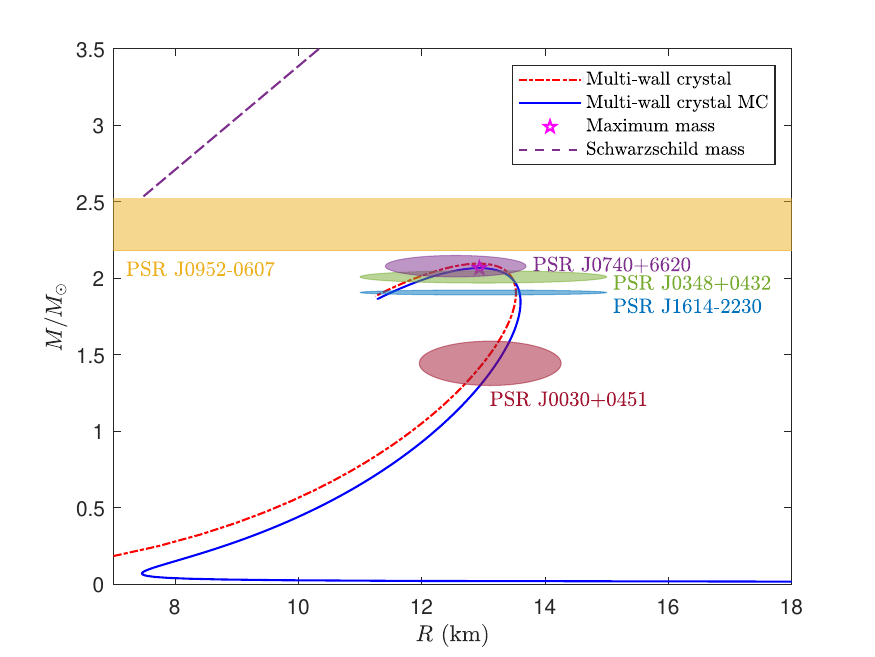}
    \caption{Mass-radius curves for neutron stars obtained from the multi-wall crystal EoS with (blue curve) and without (red curve) the Maxwell construction.
    The maximal mass $M_\textup{max}$ obtained from the MC multi-wall crystal EoS is also shown.}
    \label{fig: Plot_M-R Curve}
\end{figure}

\begin{figure}[t]
    \centering
    \includegraphics[width=0.5\textwidth]{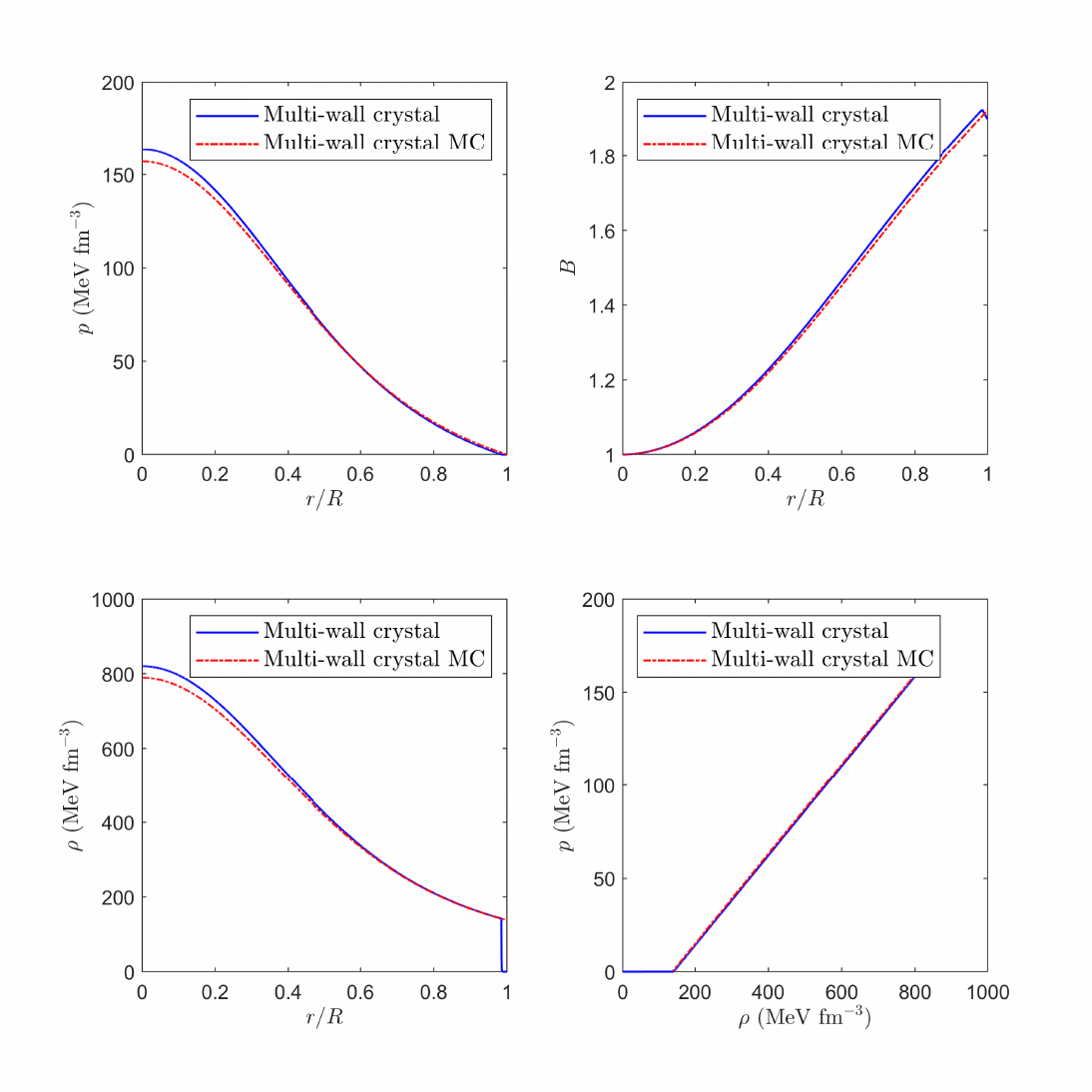}
    \caption{Plots at $M_\textup{max}=2.0971M_\odot$ of the pressure $p$, energy density $\rho$, metric function $B(r)$ and equations of state $\rho=\rho(p)$.
    The blue curve is for the crystal EoS with the Maxwell construction applied, removing any negative pressure from the system, whereas the red curve is for the ``true'' crystal EoS.}
    \label{fig: neutron star properties}
\end{figure}

Now we solve the TOV equations using the EoS obtained from the isospin asymmetric multi-wall crystal solution in the generalized $\mathcal{L}_{0246}$-Skyrme model.
In Fig. \ref{fig: Plot_M-R Curve} we present the mass-radius curve for the MC crystal (blue line) together with recent astrophysical observations.
It can be seen clearly that the obtained mass-radius curve passes through many observational constraints.
For our choice of coupling constants~\eqref{eq: Coupling constants}, the Skyrme model generates an EoS which supports rather heavy neutron stars, $M > 2M_\odot$.
Indeed, the maximum mass is predicted to be $M_\textup{max}=2.0971M_\odot$, occurring for a neutron star of radius $R=13.12\,\textup{km}$.
For this solution the central energy density is $\rho(0)=784\,\textup{MeV fm}^{-3}$, while the
central pressure is $p(0)=155.7\,\textup{MeV fm}^{-3}$.
The associated plots as a function of the maximal neutron star radius is shown in Fig.~\ref{fig: neutron star properties}.
We find that the speed of sound in the core is approximately half of the speed of light, $c_s = 0.491c$.
The maximal mass can be further increased if we assume higher value of the sextic term coupling constant $\lambda$, at the cost of increasing the corresponding radius.

The main improvement presented by the generalized multi-wall crystal, in comparison to previous studies involving the $\textup{SC}_{1/2}$ crystal, is in the low density regime.
In previous attempts, except the pure BPS Skyrme case, neutron stars obtained from Skyrme models did not have crusts, i.e. the EoS was only defined up to the nuclear saturation point $n_B \geq n_0$, and not in the low density region $n_B<n_0$.
In order to obtain a crust, the $\textup{SC}_{1/2}$ crystal EoS can be smoothly joined with an EoS that well describes the low density regime, e.g. the BCPM EoS, as in \cite{Adam_2020}.
In the resulting hybrid EoS, the high density region is still described by the $\textup{SC}_{1/2}$ crystal.
This typically increases the radius of neutron star by 1-2 km, depending on the mass of the neutron star. 
However, such a construction is not required here as the EoS from the multi-wall crystal with the Maxwell construction is valid at both high and low densities, naturally giving the neuron star a crust.


\section{Conclusion}

In the present paper, for the first time, we have obtained a ground state crystalline configuration for the generalized $\mathcal{L}_{0246}$-Skyrme model at finite densities.
In contrast to previous studies on the generalized model, it has been carried out without imposing any constraints on the geometry.
The only limiting assumption is the amount of the baryon charge hosted by the unit cell, which is $B_{\textup{cell}}=4$.
For that, we had to solve a variational problem which involves both the matter Skyrme field $\varphi$ and the metric $g$ of the unit $3$-torus $\mathbb{T}^3$.  

For our choice of the values for the coupling constants~\eqref{eq: Coupling constants}, we determine the ground state solution in the $\mathcal{L}_{0246}$-model to be the multi-wall crystal, as was recently observed by Harland \textit{et al.} \cite{Leask_2023} in the context of the $\mathcal{L}_{024}$-model.
At low densities this solution takes the form of an isolated and planar two-wall layer of skyrmionic matter.
As the baryon density grows $n_B > n_0$ then there appears to be a restoration of chiral symmetry, and the solution tends to the cubic $\textup{SC}_{1/2}$ crystal.

We have used this multi-wall crystal to investigate the three most outstanding issues of the Skyrme model in its application to dense nuclear matter and neutron stars.
Namely, (i) the problem of the thermodynamic instability at low densities; (ii) the maximal mass problem; and (iii) the compression modulus problem. 

Firstly, in comparison with the $\textup{SC}_{1/2}$ crystal or non-homogeneous crystals (e.g. $B=32$ or $B=108$ crystals composed of $\alpha$-particles), the use of the true ground state solution allowed to resolve the issue of thermodynamically instability at low densities. 
Namely, the classical energy per baryon (of the unit cell) again reveals a minimum identified with the nuclear saturation point, but now the difference between the energy at this point and at zero density is less than one percent. 
After inclusion of the quantum corrections to the total energy, due to the isospin d.o.f., and the lepton energy contributions for a $\beta$-equilibrated crystal, the total energy $E_\textup{cell}$ of the isospin asymmetric multi-wall crystal as a function of the nuclear density $n_B$ was obtained.
This minimum still existed but had reduced significantly and is practically negligible.
The energy difference used in the Maxwell construction is so small that it is difficult to tell if the minimum truly exists or if it is just an artifact of our numerical algorithm.
Nevertheless, it was still present so we had to use the Maxwell construction, which allowed us to obtain an EoS valid at all densities within the Skyrme model.

We remark that the Maxwell construction was required to avoid a thermodynamically unstable region which formally has negative pressure.
Similar regions were found in previous studies where $\alpha$, $B=32$ or $B=108$-crystals were studied.
However, it is worth underlining that in these cases the Maxwell construction was impossible (c.f. the $\alpha$-crystal) or extended to unacceptably large pressure/density regions (e.g. the corresponding neutron stars would possess cores mainly filled up by such regions).
In the current work, the pressure at which the Maxwell construction is applied is only $p_{\textup{MX}}=0.022$ MeV fm$^{-3}$ and it extends to densities below the saturation point.
Consequently, our neutron stars are mainly governed by the part of EoS above $p_{\textup{MX}}$, which is described by the multi-wall crystal EoS.

Of course, it is premature to identify the non-homogeneous low density solution found here with nuclear pasta or lasagna phases in the crust of neutron stars.
This is due to the fact that such phases emerge due to a balance between the nuclear and electrostatic forces.
However, in our study, the Coulomb interaction has not been taken into account.
In particular, we emphasize that, while our crystal qualitatively looks like nuclear pasta, it does not model nuclear pasta.
Be that as it may, our result shows that the Skyrme model itself has a tendency to form complicated, geometrically non-trivial and non-homogeneous structures at low density.
It should be again underlined that, on the contrary to all previous studies, we did not impose any geometry restrictions on the solutions, e.g. by assuming particular boundary conditions as in \cite{Canfora:2018rdz,Canfora:2020kyj}.

However, already at this stage of research, the multi-wall crystal in the density regime below saturation, $n_B<n_0$, leads to novel and intriguing observations. 
The first is the symmetry energy's disclosure of its cusp structure below the nuclear saturation density, $n_* \sim 3n_0/4<n_0$, and, secondly, the finite value of the symmetry energy in the zero density limit, $n_B \rightarrow 0$.
A cusp in the symmetry energy has previously been advocated for in \cite{Rho_2011}, wherein they attributed the presence of this cusp to a change in topology due to a transition between the FCC crystal of hedgehog skyrmions and the $\textup{SC}_{1/2}$ crystal.
A key component of their argument relies on this transition occurring in the high density regime $n_B>n_0$, however, this transition is believed to take place in the low density regime $n_B<n_0$ \cite{Adam_2022}.
However, we have argued that these two features are generic of the Skyrme model and should occur for any infinite nuclear matter that undergoes a phase transition to somewhat isolated and finite matter in the zero density limit.
This asymptotic transition to finite matter in the zero density limit is essential as the isolated solution will have a finite isospin moment of inertia tensor.
A prime example of a crystalline solution in which such a transition occurs is that of the $\alpha$-crystal, which tends to the isolated $\alpha$-particle solution as $n_B \rightarrow 0$.
Therefore, both the presence of the cusp and the non-zero value of the symmetry energy at the vacuum can be attributed as generic properties of the Skyrme model.

In fact, we have observed a further key feature of the symmetry energy.
That is, a direct correspondence between the value of the symmetry energy at the vacuum and the asymmetry energy in the Bethe–Weizs\"{a}cker SEMF for nuclear binding energies.
This strengthens our suggestion that the Skyrme model can be interpreted as a natural interpolation between infinite isospin asymmetric nuclear matter and finite (almost) symmetric atomic nuclei.
This is further supported by the observation that the proton fraction $\gamma_p \rightarrow 0.5$ in the zero density limit $n_B \rightarrow 0$, which describes almost totally isospin symmetric nuclear matter, and then, for small densities, decreases yielding asymmetric matter.
In this pattern one may again recognize finite nuclei.
Indeed, the proton number and neutron number are approximately equivalent $(\delta \approx 0)$ for smaller atomic nuclei while for larger nuclei there is an asymmetry $(\delta \neq 0)$ caused by a surplus of neutrons. 

The second big issue is also resolved since the inclusion of the sextic term makes the EoS sufficiently stiff at large densities. 
Using this EoS we were able to compute the mass-radius curve for the resulting neutron stars.
The maximal mass was found to be $M_{\textup{max}}=2.0971 M_{\odot}$, which is a acceptable large mass and the mass-radius curve fits very well to known astrophysical data. 

Finally, we shown that the problem of the compression modulus cannot be solved solely by consideration of the newly discovered non-homogeneous ground state crystalline configuration.
Although reduced by approximately $200$ MeV, the compression modulus is still a few times larger than the experimental value.
We underline that this negative result is of high importance for the Skyrme model.
It simply shows that the the solitonic model based entirely on the lightest, pionic d.o.f. is not able to correctly describe this quantity.
Fundamentally, the compression modulus is related to nuclear binding energies, which is also a problem within the Skyrme model.
If a variant of the Skyrme model has low binding energies then, naturally, the compression modulus will closer to its accepted value.
Therefore, inclusion of more massive mesons, which are known to soften the EoS at the saturation point, seems to be unavoidable.
Interestingly, this coincides with the role playing by $\rho$ mesons in reducing of the binding energies of the Skyrmions. 

It should be underlined that, if compared with other effective nuclear models, the generalized Skyrme model has an extremely small number of free parameters.
It has  only four coupling constants $\{F_\pi,m_\pi,e,\lambda\}$, of which the pion mass $m_\pi$ and the pion decay constant $F_\pi$ are, from the onset, fixed to their physical values, or as close to them as possible.
The two other parameters $e$ and $\lambda$, which, respectively, multiply the quartic (Skyrme) and sextic terms can be treated as free parameters in this model.
They can be constrained by fitting the multi-wall crystal to nuclear observables, i.e. they can be chosen such that the symmetric energy $M_B(n_B)$ and nuclear density $n_B$ at saturation $n_0$ are close to the experimentally determined values. 

There are several directions in which our study can be continued. 
First of all, it is widely known that the lower density phases of nuclear matter are governed by a balance between nuclear and Coulomb forces, which leads to a plethora of geometrically different structures.
The fact that the generalized Skyrme model, even without the inclusion of electrostatic interactions, gives rise to the multi-wall crystal (a lasagna like structure) can be viewed as an intrinsic ability of the model to provide such solutions.
Other non-homogeneous configurations have been observed in the Skyrme model \cite{Canfora:2018rdz,Canfora:2020kyj}, however they were an effect of the imposed boundary conditions and therefore their applications to nuclear physics remain to be clarified.
Undoubtedly, inclusion of the Coulomb interaction seems mandatory, see e.g. \cite{Ma:2019fvk}.
It seems likely that including Coulomb interactions will not only give insight into such geometric phases but could also allow one to avoid use of the Maxwell construction.
Thus it could possibly provide a complete description of the crust in neutron star within the Skyrme model framework. 

More importantly, the inclusion of other d.o.f., like for example $\rho$ or $\omega$ mesons, seems inevitable to resolve the issue of the compressibility at nuclear saturation.
This, combined with the inhomogeneous multi-wall crystal detailed in the paper, may possibly lead to the correct value of the compression modulus.


\section*{Acknowledgments}
PL is supported by a Ph.D. studentship from UKRI, Grant No. EP/V520081/1. MHG thanks the Xunta de Galicia (Consellería de Cultura, Educación y Universidad) for the funding of his predoctoral activity through Programa de ayudas a la etapa predoctoral 2021. AW was supported by the Polish National Science center (NCN 2020/39/B/ST2/01553).
The authors thanks Christoph Adam and Alberto Garcia Martin-Caro for discussions and comments.


\appendix


\section{Extended virial constraints}

The space of allowed variations $\mathscr{E}$ is a $6$-dimensional subspace of the space of sections of the rank $6$ vector bundle $\odot^2 T^*\mathbb{T}^3$,
\begin{equation}
    \mathscr{E} = \left\{ \delta g_{ij} \textup{d}x^i \textup{d}x^j \in \Gamma(\odot^2 T^*\mathbb{T}^3) : \delta g_{ij} \, \textup{constant} \right\}.
\end{equation}
By definition, the energy $M_B$ is critical with respect to variations $g_s$ of the metric if and only if
\begin{equation}
    \left.\frac{\textup{d} M_B(\varphi, g_s)}{\textup{d}s}\right|_{s=0} = \int_{\mathbb{T}^3} \textup{d}^3x \sqrt{g} \braket{S(\varphi,g), \delta g}_g = 0,
\end{equation}
that is, if and only if $S \perp_{L^2} \mathscr{E}$.
Now let the orthogonal compliment of $g$ in $\mathscr{E}$, the space of traceless parallel symmetric bilinear forms, given by
\begin{equation}
    \mathscr{E}_0 = \left\{ \theta \in \Gamma(\odot^2 T^*\mathbb{T}^3) : \Tr_g(\theta) = \braket{\theta,g}_g = 0 \right\}.
\end{equation}
Then the criticality condition $S \perp_{L^2} \mathscr{E}$ can be reformulated as \cite{Speight_2014}
\begin{equation}
    \int_{\mathbb{T}^3} \textup{d}^3x \sqrt{g} \braket{S(\varphi,g), g}_g = 0 \quad \textup{and} \quad S \perp_{L^2} \mathscr{E}_0.
\end{equation}
The first condition $S \perp_{L^2} g$ is analogous to a virial constraint and the second condition $S \perp_{L^2} \mathscr{E}_0$ coincides with the extended virial constraints derived by Manton \cite{Manton_2009}.
We can determine the virial constraint by using the trace~\eqref{eq: Trace} and evaluating
\begin{align}
    \int_{\mathbb{T}^3} \braket{S(\varphi,g), g}_g \textup{vol}_g = \, & \int_{\mathbb{T}^3} \textup{d}^3x \sqrt{g} \Tr_g(S) \nonumber \\
    = \, & \frac{1}{2} \left( E_2 - E_4 + 3 E_0 - 3 E_6 \right).
\end{align}
Hence, the condition $S \perp_{L^2} g$ establishes the familiar virial constraint
\begin{equation}
    E_2 - E_4 + 3 (E_0 - E_6) = 0.
\end{equation}

To determine the extended virial constraint corresponding to the condition $S \perp_{L^2} \mathscr{E}_0$, we define a symmetric bilinear form $\Delta: T_x \mathbb{T}^3 \times T_x \mathbb{T}^3 \rightarrow \mathbb{R}$,
\begin{equation}
    \begin{split}
        \Delta_{ij} = -\int_{\mathbb{T}^3} \textup{d}^3x \sqrt{g} \left( \frac{c_2}{2} \Tr(L_i L_j) \right. \\ \left. + \frac{c_4}{2}g^{kl}\Tr([L_i,L_k][L_j,L_l]) \right).
    \end{split}
\end{equation}
In the metric independent integral formulation, this symmetric bilinear form $\Delta$ reads
\begin{align}
    \Delta_{ij} = \sqrt{g} \, L_{ij}(\varphi) + 2 \sqrt{g} \, g^{kl} \Omega_{ikjl}(\varphi).
\end{align}
Then $S \perp_{L^2} \mathscr{E}_0$ if and only if $\Delta$ is orthogonal to $\mathscr{E}_0$ with respect to the inner product $\braket{\cdot,\cdot}_{\mathscr{E}}$.
Therefore, for $\lambda \in \mathbb{R}$ we must have
\begin{equation}
    \Delta = \lambda g.
\end{equation}
Taking the trace of both sides yields
\begin{align}
    3 \lambda = E_2 + 2E_4.
\end{align}
Thus, the condition $S \perp_{L^2} \mathscr{E}_0$ produces the extended virial constraint
\begin{equation}
    \Delta = \frac{1}{3} \left( E_2 + 2E_4 \right)g.
\end{equation}
So we see that $\varphi: \mathbb{T}^3 \rightarrow \SU(2)$ is a skyrmion crystal if and only if it satisfies the extended virial constraints:
\begin{subequations}
    \begin{align}
        &E_2 - E_4 = 3 (E_6 - E_0), \\
        &\Delta = \frac{1}{3} \left( E_2 + 2E_4 \right)g.
    \end{align}
\label{eq: Extended virial constraints}
\end{subequations}
We will verify numerically that the extended virial constraints are being satisfied within some tolerance, e.g. $\textup{tol} = 10^{-5}$.
This is done by checking that
\begin{equation}
    \left|\frac{E_4}{E_2+ 3(E_0-E_6)} - 1 \right| < \textup{tol}
\end{equation}
and
\begin{equation}
    \left|\frac{\Delta_{ij}}{(E_4+E_6-E_0)g_{ij}} - 1\right| < \textup{tol}.
\end{equation}


\section{Reconstructing $\Lambda$ from $g$}

As the metric $g_s$ on $\mathbb{T}^3$ varies so too does the lattice $\Lambda_s$, which we have labeled $\Lambda_s = \Lambda(g_s)$ where $\Lambda_0 = \Lambda$.
As before, let $\Lambda_\diamond$ be the energy minimising lattice and denote the corresponding energy minimising metric on $\mathbb{T}^3$ by $g_{\diamond}$.
Let $\vec{X}_1=(x_1,y_1,z_1), \vec{X}_2=(x_2,y_2,z_2)$ and $\vec{X}_3=(x_3,y_3,z_3)$ be the period lattice vectors for $\Lambda_\diamond$.
In order to plot isosurfaces of the baryon density of the resulting skyrmion on $(\mathbb{R}^3/\Lambda_\diamond,d)$, we need to reconstruct the lattice $\Lambda_\diamond$ from the metric $g_{\diamond}$.
To do this we need to solve the following under-determined system of equations
\begin{equation}
    \begin{matrix}
        \vec{X}_1 \cdot \vec{X}_1 = x_1^2 + y_1^2 + z_1^2 = g_{11} \\
        \vec{X}_1 \cdot \vec{X}_2 = x_1 x_2 + y_1 y_2 + z_1 z_2 = g_{12} \\
        \vec{X}_1 \cdot \vec{X}_3 = x_1 x_3 + y_1 y_3 + z_1 z_3 = g_{13} \\
        \vec{X}_2 \cdot \vec{X}_2 = x_2^2 + y_2^2 + z_2^2 = g_{22} \\
        \vec{X}_2 \cdot \vec{X}_3 = x_2 x_3 + y_2 y_3 + z_2 z_3 = g_{23} \\
        \vec{X}_3 \cdot \vec{X}_3 = x_3^2 + y_3^2 + z_3^2 = g_{33}
    \end{matrix},
\label{eq: Skyrme crystals - Underdetermined system of equations}
\end{equation}
where we have written $g_{ij}=(g_{\diamond})_{ij}$ for notational convenience.
This has infinitely many solutions which we can solve for by fixing a particular lattice vector, or by setting $y_1=z_1=z_2=0$, i.e. $\vec{X}_1=(x_1,0,0), \vec{X}_2=(x_2,y_2,0)$ and $\vec{X}_3=(x_3,y_3,z_3)$.
Then, for the latter choice of period lattice vectors, the system of equations~\eqref{eq: Skyrme crystals - Underdetermined system of equations} has a unique solution given by
\begin{equation}
    \vec{X}_1 = \left( \sqrt{g_{11}}, 0, 0 \right),
\end{equation}
\begin{equation}
    \vec{X}_2 = \left( \frac{g_{12}}{\sqrt{g_{11}}}, \sqrt{g_{22} - \frac{g_{12}^2}{g_{11}}}, 0 \right),
\end{equation}
\begin{equation}
    \begin{split}
        \vec{X}_3 = \left( \frac{g_{13}}{\sqrt{g_{11}}}, \frac{1}{\sqrt{g_{22}-\frac{g_{12}^2}{g_{11}}}} \left( g_{23} - \frac{g_{12}g_{13}}{g_{11}} \right), \right. \\ \left.
        \sqrt{g_{33} - \frac{g_{13}^2}{g_{11}} - \frac{1}{\left(g_{22}-\frac{g_{12}^2}{g_{11}}\right)} \left( g_{23} - \frac{g_{12}g_{13}}{g_{11}} \right)^2} \right).
    \end{split}
\end{equation}


\section{Derivation of the isospin inertia tensor in $\sigma$-model notation}

Under the dynamical transformation \eqref{eq: RBQ - Dynamical ansatz}, the Dirichlet energy transforms as
\begin{align}
    \mathcal{L}_2 = \, & \frac{c_2}{2} g^{\mu\nu}\Tr\left(\hat{L}_{\mu}\hat{L}_{\nu}\right) \nonumber \\
    = \, & \frac{c_2}{2} g^{00} \Tr\left(\hat{L}_{0}\hat{L}_{0}\right) + \frac{c_2}{2} g^{ij} \Tr\left(\hat{L}_{i}\hat{L}_{j}\right) \nonumber \\
    = \, & -\frac{c_2}{2} \Tr\left(T_i T_j\right) \omega_i \omega_j + \frac{c_2}{2} g^{ij} \Tr\left(L_i L_j\right),
\end{align}
where the first term is the Dirichlet energy contribution to the isospin inertia tensor, and the second term is the static Dirichlet energy.
Likewise, for the Skyrme term we have
\begin{align}
    \mathcal{L}_4 = \, & \frac{c_4}{4} g^{\mu\alpha}g^{\nu\beta} \Tr\left([\hat{L}_{\mu},\hat{L}_{\nu}][\hat{L}_{\alpha},\hat{L}_{\beta}]\right) \nonumber \\
    = \, & \frac{c_4}{2} g^{00} g^{kl} \Tr\left([\hat{L}_{0},\hat{L}_{k}][\hat{L}_{0}\hat{L}_{l}]\right) \nonumber \\
    \, & + \frac{c_4}{4}g^{ik}g^{jl} \Tr\left([\hat{L}_{i},\hat{L}_{j}][\hat{L}_{k},\hat{L}_{l}]\right) \nonumber \\
     = \, & -\frac{c_4}{2} g^{kl} \Tr\left([T_i,L_k][T_j,L_l]\right) \omega_i \omega_j \nonumber \\
     \, & + \frac{c_4}{4} g^{ik}g^{jl} \Tr\left([L_i,L_j][L_k,L_l]\right),
\end{align}
where the first term is the Skyrme contribution to the isospin inertia tensor.
Finally, the sextic term,
\begin{align}
    \mathcal{L}_6 = \, & -c_6 g^{\mu\nu} \frac{\epsilon^{\mu\alpha\beta\gamma}\epsilon^{\nu\delta\rho\sigma} }{(24 \pi^2 \sqrt{-g})^2} \Tr(\hat{L}_\alpha \hat{L}_\beta \hat{L}_\gamma) \Tr(\hat{L}_\delta \hat{L}_\rho \hat{L}_\sigma) \nonumber \\
    = \, & -c_6 g^{00} \frac{\epsilon^{ijk}\epsilon^{abc} }{(24 \pi^2 \sqrt{-g})^2} \Tr(L_i L_j L_k) \Tr(L_a L_b L_c) \nonumber \\
    \, & -c_6 g^{kl} \frac{3^2 \epsilon^{kab}\epsilon^{lcd} }{(24 \pi^2 \sqrt{-g})^2} \Tr(\hat{L}_0 \hat{L}_a \hat{L}_b) \Tr(\hat{L}_0 \hat{L}_c \hat{L}_d) \nonumber \\
     = \, &  - c_6 g^{kl} \frac{\epsilon^{kab}\epsilon^{lcd}}{(8 \pi^2 \sqrt{-g})^2} \Tr(T_i L_a L_b) \Tr(T_j L_c L_d) \omega_i \omega_j \nonumber \\
     \, & + c_6 \left(\mathcal{B}^0\right)^2.
\end{align}
As the static part of the sextic term is the temporal component, we must take the negative contribution of this.
Putting all of this together, we find the effective Lagrangian to be
\begin{align}
    \mathcal{L}_{\textup{eff}} = \, & \mathcal{L}_0 + \mathcal{L}_2 + \mathcal{L}_4 - \mathcal{L}_6 \nonumber \\
    = \, & -c_0 M_\pi^2 \Tr\left( \Id - \varphi \right) + \frac{c_2}{2} g^{ij} \Tr\left(L_i L_j\right) \nonumber \\
    \, & + \frac{c_4}{4} g^{ik}g^{jl} \Tr\left([L_i,L_j][L_k,L_l]\right) - c_6 \left(\mathcal{B}^0\right)^2 \nonumber \\
    & + \frac{1}{2}\left\{ -c_2\Tr\left(T_i T_j\right) -c_4 g^{kl} \Tr\left([T_i,L_k][T_j,L_l]\right) \right. \nonumber \\
    \, & \left. + c_6 g^{kl} \frac{\epsilon^{kab}\epsilon^{lcd}}{(4\sqrt{2} \pi^2 \sqrt{-g})^2} \Tr(T_i L_a L_b) \Tr(T_j L_c L_d) \right\} \omega_i \omega_j \nonumber \\
    = \, & -\mathcal{E}_\textup{stat} + \frac{1}{2}\omega_i \mathcal{U}_{ij} \omega_j,
\end{align}
where the isospin inertia tensor density contribution from the Skyrme field $\varphi$ is given by \eqref{eq: Isospin inertia tensor density}.

In the quaternionic formulation, the $\su(2)$ current $T_i$ is expressed by the vector quaternion
\begin{equation}
    T_i = -i T_i^a \tau^a, \quad T_i^j = \delta^{ij}\varphi^k\varphi^k - \varphi^i\varphi^j - \epsilon^{ijk}\varphi^0\varphi^k.
\end{equation}
The corresponding contractions are found to be
\begin{subequations}
    \begin{align}
        T_i^k T_j^k = \, & \delta^{ij}\varphi^k\varphi^k - \varphi^i\varphi^j, \\
        T_i^k L_j^k = \, & -\epsilon^{ikl} \varphi^k \partial_j \varphi^l.
    \end{align}
\end{subequations}
Therefore, the Dirichlet contribution can be written as
\begin{equation}
    \Tr(T_i T_j) = -2 \left\{ \delta^{ij}\varphi^k\varphi^k - \varphi^i\varphi^j \right\}.
\end{equation}
After a painstaking, but straightforward, calculation one finds that
\begin{equation}
    \begin{split}
        \Tr\left([T_i, L_a][T_j, L_b]\right) = -8 \left\{(\delta^{ij}- \varphi^i\varphi^j)\partial_a\varphi^0\partial_b\varphi^0 \right. \\ \left. + (\varphi^c\varphi^c)\partial_a\varphi^i\partial_b\varphi^j + \varphi^0\varphi^i\partial_a\varphi^0\partial_b\varphi^j + \varphi^0\varphi^j\partial_b\varphi^0\partial_a\varphi^i \right\}.
    \end{split}
\end{equation}
Then, finally, we need to consider the term
\begin{equation}
    \begin{split}
        \Tr(T_i L_m L_n)\Tr(T_j L_k L_l) = 4 \epsilon_{pqr} \epsilon_{cde} T_i^p L_m^q L_n^r T_j^c L_k^d L_l^e  \\
        = 4 T_i^p T_j^p \left( L_m^q L_k^q L_n^r L_l^r - L_n^r L_k^r L_m^q L_l^q \right) \\
        + 4 T_j^p L_m^p \left( T_i^q L_l^q L_n^r L_k^r - T_i^q L_k^q L_n^r L_l^r \right) \\
        + 4 T_j^p L_n^p \left( T_i^q L_k^q L_m^r L_l^r - T_i^q L_l^q L_m^r L_k^r \right).
    \end{split}
\end{equation}
Putting all of this together by using the quaternion representation~\eqref{eq: Quaternion representation}, the isospin inertia tensor density takes the form
\begin{widetext}
    \begin{align}
    \mathcal{U}_{ij} = \, & 2c_2\left( \delta^{ij}\varphi^k\varphi^k - \varphi^i\varphi^j \right) + 8 c_4 g^{kl} \left((\delta^{ij}- \varphi^i\varphi^j)\partial_k\varphi^0\partial_l\varphi^0 + (\varphi^m\varphi^m)\partial_k\varphi^i\partial_l\varphi^j + \varphi^0\varphi^i\partial_k\varphi^0\partial_l\varphi^j + \varphi^0\varphi^j\partial_l\varphi^0\partial_k\varphi^i \right)  \nonumber \\
    & + \frac{2c_6}{(4\pi^2 \sqrt{-g})^2} g_{pq} \epsilon^{pmn}\epsilon^{qkl} \left[ (\delta^{ij}\varphi^a\varphi^a - \varphi^i\varphi^j) \left(\partial_m\varphi^\mu \partial_k\varphi^\mu \partial_n\varphi^\nu \partial_l\varphi^\nu - \partial_n\varphi^\mu \partial_k\varphi^\mu \partial_m\varphi^\nu \partial_l\varphi^\nu\right) \right. \nonumber \\
    & + \epsilon^{jac} \varphi^a \partial_m \varphi^c \left( \epsilon^{ibd} \varphi^b \partial_l \varphi^d \partial_n\varphi^\mu \partial_k\varphi^\mu - \epsilon^{ibd} \varphi^b \partial_k \varphi^d \partial_n\varphi^\mu \partial_l\varphi^\mu \right) \nonumber \\
    & \left. + \epsilon^{jac} \varphi^a \partial_n \varphi^c \left( \epsilon^{ibd} \varphi^b \partial_k \varphi^d \partial_m\varphi^\mu \partial_l\varphi^\mu - \epsilon^{ibd} \varphi^b \partial_l \varphi^d \partial_m\varphi^\mu \partial_k\varphi^\mu \right) \right].
\end{align}
\end{widetext}


\section{The Tolman--Oppenheimer--Volkoff equations}

From the metric ansatz \eqref{eq: Spherical metric ansatz}, we can determine the Christoffel symbols
\begin{equation}
    \Gamma_{\mu\nu}^\lambda = \frac{1}{2} g^{\lambda\sigma}\left( \partial_\mu g_{\nu\sigma} + \partial_\nu g_{\mu\sigma} - \partial_\sigma g_{\mu\nu} \right),
\end{equation}
of which the non-zero components are found to be
\begin{equation}
    \begin{split}
        \Gamma_{tt}^r=\Gamma_{tr}^t = \frac{1}{2A}\frac{\textup{d}A}{\textup{d}r},
        \quad \Gamma_{rt}^t = \frac{1}{2B}\frac{\textup{d}A}{\textup{d}r}, \\
        \Gamma_{rr}^r = \frac{1}{2B}\frac{\textup{d}B}{\textup{d}r},
        \quad \Gamma_{\phi\theta}^\phi = \Gamma_{\phi\phi}^\theta = \cot\theta, \\
        \Gamma_{r\theta}^\theta = -\frac{r}{B}, \quad \Gamma_{\theta r}^\theta = \Gamma_{\theta\theta}^r = \Gamma_{\phi r}^\phi = \Gamma_{\phi\phi}^r = \frac{1}{r}, \\
        \Gamma_{r\phi}^\phi = -\frac{r\sin^2\theta}{B}, \quad \Gamma_{\theta\phi}^\phi = -\sin\theta\cos\theta. 
    \end{split}
\label{eq: Christoffel symbols}
\end{equation}
Thus the Riemann curvature tensor can be obtained using the non-zero Christoffel symbols~\eqref{eq: Christoffel symbols},
\begin{equation}
    R^\sigma_{\rho\mu\nu} = \partial_\mu \Gamma_{\nu\rho}^\sigma - \partial_\nu \Gamma_{\mu\rho}^\sigma + \Gamma_{\nu\rho}^\lambda \Gamma_{\mu\lambda}^\sigma - \Gamma_{\mu\rho}^\lambda \Gamma_{\nu\lambda}^\sigma.
\end{equation}
The Ricci tensor is given by $R_{\mu\nu}=g^{\rho\sigma}R_{\rho\mu\sigma\nu}$ and the relevant components are found to be given by
\begin{equation}
    \begin{split}
        R_{tt} = -\frac{1}{4B^2} \left[ \frac{\textup{d}A}{\textup{d}r}\frac{\textup{d}B}{\textup{d}r} + B\left( - \frac{4}{r}\frac{\textup{d}A}{\textup{d}r} \right.\right. \\ \left.\left. + \frac{1}{A}\left(\frac{\textup{d}A}{\textup{d}r}\right)^2 - 2\frac{\textup{d}^2A}{\textup{d}r^2} \right) \right]
    \end{split}
\end{equation}
and
\begin{equation}
    \begin{split}
        R_{rr} = \frac{1}{4A^2 B r} \left[A \frac{\textup{d}B}{\textup{d}r}\left( 4A + r\frac{\textup{d}A}{\textup{d}r} \right) \right. \\  \left. + Br\left( \left( \frac{\textup{d}A}{\textup{d}r} \right)^2 - 2A \frac{\textup{d}^2A}{\textup{d}r^2}\right) \right]. 
    \end{split}
\end{equation}
Now we can compute the Ricci scalar $R=g^{\mu\nu}R_{\mu\nu}$, that is
\begin{equation}
    \begin{split}
        R = \frac{1}{2A^2B^2r^2} \left[ Br^2\left(\frac{\textup{d}A}{\textup{d}r}\right)^2 + 4A^2\left( r\frac{\textup{d}B}{\textup{d}r} + B^2 - B\right)  \right. \\ \left. +Ar \left( r \frac{\textup{d}A}{\textup{d}r}\frac{\textup{d}B}{\textup{d}r} -2B\left( r\frac{\textup{d}^2A}{\textup{d}r^2} + 2\frac{\textup{d}A}{\textup{d}r}\right) \right) \right].
    \end{split}
\end{equation}
Now we have all the ingredients required to compute the Einstein tensor, $G_{\mu\nu}=R_{\mu\nu}-\frac{1}{2}Rg_{\mu\nu}$.
The relevant components of the Einstein tensor are found to be
\begin{subequations}
    \begin{align}
        G_{tt} = \, & \frac{A(r)}{B(r)^2 r^2} \left[ r\frac{\textup{d}B(r)}{\textup{d}r} + B(r)\left( B(r) -1\right)\right], \\
        G_{rr} = \, & \frac{1}{A(r) r^2} \left[ r\frac{\textup{d}A(r)}{\textup{d}r} - A(r)\left( B(r) -1\right) \right].
    \end{align}
\label{eq: Einstein tensor}
\end{subequations}

In the static case, and for a diagonal metric (that of which is applicable to us), we have $u_\mu=(\sqrt{-g_{00}},0,0,0)$ and the non-zero components of the energy-momentum tensor are given by
\begin{equation}
    T_{00} = -\rho(p(r)) g_{00}, \quad T_{ij} = p(r) g_{ij}.
\end{equation}
In particular, for the spherical metric ansatz \eqref{eq: Spherical metric ansatz}, the energy-momentum tensor reduces to the four terms:
\begin{subequations}
    \begin{align}
        T_{tt} = \, & \rho(p(r)) A(r), \\
        T_{rr} = \, & B(r) p(r), \\
        T_{\theta\theta} = \, & r^2 p(r), \\
        T_{\phi\phi} = \, & r^2 p(r) \sin^2\theta.
    \end{align}
\label{eq: Energy-momentum tensor}
\end{subequations}
We are now in a position to calculate the Einstein equations~\eqref{eq: Einstein equation} by using the energy-momentum tensor~\eqref{eq: Energy-momentum tensor} and the Einstein tensor~\eqref{eq: Einstein tensor}.
From this, and the Bianchi identity
\begin{equation}
    0 = \nabla_\nu T^{r\nu} = \frac{\partial T^{r\nu}}{\partial x^\nu} + T^{\sigma\nu} \Gamma_{\sigma\nu}^r + T^{r\sigma} \Gamma_{\sigma\nu}^\nu,
\end{equation}
we get the TOV system of ODEs \eqref{eq: TOV system}.


\bibliography{main}

\end{document}